\documentclass[a4paper,11pt]{article}
\usepackage{heppub}
\pdfoutput=1
\usepackage{graphicx}

\usepackage{subcaption}
\usepackage{placeins}

\newcommand{\Abs}[1]{\bigl\lvert#1\bigr\rvert}
\newcommand{\ord}[1]{\mathcal{O}(#1)}

\newcommand{\df}{\mathrm{d}}
\newcommand{\eps}{\epsilon}
\newcommand{\dY}{\Delta y}

\newcommand{\cO}{{\mathcal O}}

\newcommand{\cM}{{\mathcal M}}
\newcommand{\Tau}{{\mathcal T}}

\newcommand{\bn}{{\bar{n}}}
\newcommand{\gamgam}{{\gamma\gamma}}

\newcommand{\GeV}{\,\mathrm{GeV}}

\newcommand{\nn}{\nonumber}
\renewcommand{\min}{{\rm min}}

\newcommand{\Msquared}{A}
\newcommand{\as}{\alpha_s}

\newcommand{\cut}{{\mathrm{cut}}}
\newcommand{\cuts}{{(\mathrm{cuts})}}
\newcommand{\Ecm}{E_\mathrm{cm}}

\newcommand{\sub}{\mathrm{sub}}
\newcommand{\off}{\mathrm{off}}
\newcommand{\LO}{\mathrm{LO}}

\newcommand{\lep}{\mathrm{lep}}
\newcommand{\iso}{{\rm iso}}
\newcommand{\ETiso}{E_T^\iso}
\newcommand{\zero}{{(0)}}

\newcommand{\qt}{{\vec q}_T}
\newcommand{\pt}{{\vec p}_T}
\newcommand{\kt}{{\vec k}_T}

\begin{document}

%%%%%%%%%%%%%%%%%%%%%%%%%%%%%%%%%%%%%%%%%%%%%%%%%%%%%%%%%%%%%%%%%%%%%%%%%%%%%%%%
% Title page
%%%%%%%%%%%%%%%%%%%%%%%%%%%%%%%%%%%%%%%%%%%%%%%%%%%%%%%%%%%%%%%%%%%%%%%%%%%%%%%%

\title{Impact of Isolation and Fiducial Cuts on $q_T$ and N-Jettiness Subtractions}

\author[a]{Markus A.~Ebert}
\emailAdd{ebert@mit.edu}

\author[b]{and Frank J.~Tackmann}
\emailAdd{frank.tackmann@desy.de}

\affiliation[a]{Center for Theoretical Physics, Massachusetts Institute of Technology, Cambridge, MA 02139, USA}
\affiliation[b]{Theory Group, Deutsches Elektronen-Synchrotron (DESY), D-22607 Hamburg, Germany}

%%%%%%%%%%%%%%%%%%%%%%%%%%%%%%%%%%%%%%%%%%%%%%%%%%%%%%%%%%%%%%%%%%%%%%%%%%%%%%%%
\abstract
{
Kinematic selection cuts and isolation requirements are a necessity in
experimental measurements for identifying prompt leptons and photons that
originate from the hard-interaction process of interest. We analyze how such
cuts affect the application of the $q_T$ and $N$-jettiness subtraction methods for
fixed-order calculations. We consider both fixed-cone and smooth-cone isolation
methods. We find that kinematic selection and isolation cuts both induce parametrically
enhanced power corrections with considerably slower convergence compared to the
standard power corrections that are already present in inclusive cross sections
without additional cuts. Using analytic arguments at next-to-leading order we
derive their general scaling behavior as a function of the subtraction cutoff.
We also study their numerical impact for the case of gluon-fusion Higgs
production in the $H\to\gamma\gamma$ decay mode and for $pp\to\gamma\gamma$ direct
diphoton production. We find that the relative enhancement of the additional
cut-induced power corrections tends to be more severe for $q_T$, where it can
reach an order of magnitude or more, depending on the choice of parameters and
subtraction cutoffs. We discuss how all such cuts can be incorporated without
causing additional power corrections by implementing the subtractions
differentially rather than through a global slicing method. We also highlight
the close relation of this formulation of the subtractions to the
projection-to-Born method.
}
%%%%%%%%%%%%%%%%%%%%%%%%%%%%%%%%%%%%%%%%%%%%%%%%%%%%%%%%%%%%%%%%%%%%%%%%%%%%%%%%

\date{November 19, 2019}

\preprint{\vbox{%
\hbox{DESY 19-199}
\hbox{MIT--CTP/5158}
}}

\maketitle

%%%%%%%%%%%%%%%%%%%%%%%%%%%%%%%%%%%%%%%%%%%%%%%%%%%%%%%%%%%%%%%%%%%%%%%%%%%%%%%%
\section{Introduction}
\label{sec:intro}
%%%%%%%%%%%%%%%%%%%%%%%%%%%%%%%%%%%%%%%%%%%%%%%%%%%%%%%%%%%%%%%%%%%%%%%%%%%%%%%%

An important class of measurements at colliders such as the LHC
are processes involving leptons or photons in the final state.
For example, the cleanest channels to measure Higgs production
are the $H \to 4\ell$ and $H \to \gamgam$ decay modes,
and both have been studied extensively by ATLAS and CMS~\cite{Aaboud:2017oem,
Aaboud:2017vzb, Aaboud:2018xdt, Aaboud:2018wps,
Sirunyan:2017exp, Sirunyan:2018ouh, Sirunyan:2018kta, CMS-PAS-HIG-19-004}.
Other important examples are inclusive $W\to\ell\nu$ and $Z/\gamma^*\to\ell\ell$
production~\cite{Aad:2015auj, Aaboud:2017svj, Aaboud:2017ffb, CMS:2014jea,
Khachatryan:2016nbe, Sirunyan:2019bzr},
direct diphoton production $pp \to \gamgam$~\cite{Chatrchyan:2011qt, Chatrchyan:2014fsa, Aad:2012tba, Aaboud:2017vol}, and more generally any process involving prompt photons or electroweak
bosons in leptonic decay channels.
In all such measurements, lepton and photon kinematic selection cuts and
isolation requirements are necessary to identify the leptons and photons originating
from the hard interaction and to suppress backgrounds such as misidentified jets or
secondary leptons and photons arising for example from the decay of hadrons.

The most prominent selection cuts are minimum $p_T$ requirements.
The isolation is commonly achieved by restricting the energy in a cone
around the lepton or photon to be bounded, for example
\begin{align} \label{eq:iso_intro}
 \sum_{i:\, d(i,\gamma/\ell) < R} E_T^i < \ETiso
\,,\end{align}
where the sum runs over all particles $i$ in a cone of size $R$ around the photon $\gamma$
or lepton $\ell$.

Since isolation requirements as in \eq{iso_intro} are sensitive to the momenta of
all hadrons in an event, incorporating them into higher-order calculations requires
one to explicitly take into account the isolation cuts
when integrating over the phase space of real emissions. This in turn requires
fixed-order calculations that are fully exclusive in the final state of real emissions.
A key challenge in higher-order calculations is the cancellation of infrared (IR) divergences
from the soft and collinear limits of real emissions against corresponding divergences
from virtual corrections.
At NLO, fully-exclusive calculations are achieved by applying
local subtraction techniques such as the FKS~\cite{Frixione:1995ms,Frixione:1997np}
or CS~\cite{Catani:1996jh, Catani:1996vz, Catani:2002hc} subtractions.
At next-to-next-to-leading order (NNLO), local subtraction techniques become much more involved
due to the overlap of virtual and real divergences, and
a variety of such methods have been developed~\cite{GehrmannDeRidder:2005cm,
Currie:2013vh, Czakon:2010td, Czakon:2014oma, Boughezal:2011jf,
DelDuca:2015zqa, DelDuca:2016csb, Cacciari:2015jma,
Caola:2017dug, Caola:2019nzf, Herzog:2018ily, Magnea:2018hab, Magnea:2018ebr}.

Another approach to obtain fully-exclusive NNLO calculations is the use of global slicing methods
\cite{Catani:2007vq, Boughezal:2015dva, Gaunt:2015pea},
where one exploits that the cancellation of IR divergences occurs in the singular limit
of a suitable resolution variable, and that this singular limit can be predicted from a
factorization theorem. For the transverse momentum $q_T$, the relevant factorization was
first shown in \refscite{Collins:1981uk,Collins:1981va,Collins:1984kg}.
For $N$-jettiness $\Tau_N$~\cite{Stewart:2010tn}, the relevant factorization was derived in
\refscite{Stewart:2009yx, Stewart:2010tn} using the soft-collinear effective theory (SCET)
\cite{Bauer:2000ew, Bauer:2000yr, Bauer:2001ct, Bauer:2001yt, Bauer:2002nz}.
All contributions to the cross section not described by the factorization,
usually referred to as nonsingular terms or power corrections, can then be obtained
from an NLO calculation. Hence, an advantage of the slicing methods is that they are
comparably straightforward to implement, since they allow reusing much of the existing
NLO calculations. For the same reasons, they are also extendable to
N$^3$LO~\cite{Gaunt:2015pea, Cieri:2018oms, Billis:2019vxg}.

An important aspect of slicing methods is that they require a resolution cutoff,
which induces power corrections from contributions below the cutoff that are neglected.
To improve the numerical performance, these power corrections can be included
systematically by computing them in an expansion in the resolution variable
about the soft and collinear limits. Recently, there has been significant interest
and progress in understanding collider cross sections at subleading power~\cite{Laenen:2008ux, Laenen:2010uz, Bonocore:2014wua, Bonocore:2015esa, Kolodrubetz:2016uim, Bonocore:2016awd, Moult:2017rpl, Feige:2017zci, DelDuca:2017twk, Chang:2017atu, Beneke:2017ztn, Moult:2018jjd, Bahjat-Abbas:2018hpv, Beneke:2018rbh, Beneke:2018gvs, Bhattacharya:2018vph, Beneke:2019mua, Moult:2019uhz}.
In particular, for inclusive Higgs and Drell-Yan production the leading-logarithmic (LL) corrections
at NNLO at next-to-leading power (NLP) are known for $\Tau_0$~\cite{Moult:2016fqy, Boughezal:2016zws, Moult:2017jsg}. At NLO, the full NLP corrections are known for
$\Tau_0$~\cite{Boughezal:2018mvf, Ebert:2018lzn}, $q_T$~\cite{Ebert:2018gsn},
and $\Tau_1$~\cite{Boughezal:2019ggi}.

The same power corrections are also important for the resummation of logarithms $\ln(q_T/Q)$
or $\ln(\Tau_0/Q)$ in the $q_T$ or $\Tau_0$ spectra at small $q_T \ll Q$ or $\Tau_0 \ll Q$
(with $Q$ being the relevant hard-interaction scale).
This resummation is based on the same factorization theorems underlying the subtraction methods,
as the logarithmic terms precisely arise in the singular limit of the cross section.
In addition to the resummed singular cross section, one has to include the power corrections
in order to recover the full fixed-order result for the spectrum.
Thus, understanding the effect of selection and isolation cuts on the factorization
is equally important for resummation.

So far, studies of power corrections have only considered
inclusive processes, while the effect
of selection and isolation cuts have not yet been considered.
As we will see, these cuts are an additional source of power corrections.
Given their necessity for experimental measurements, it is important to study
the cut-induced power corrections, and in particular determine if and when they lead
to the dominant corrections or if they can even lead to a breakdown of the
factorization and thus the subtraction methods.

In this paper, we study the effect of kinematic selection and isolation cuts
on $q_T$ and $\Tau_0$ factorization. For concreteness, we focus on the case of diphoton production,
either in the direct process $pp \to \gamgam$ or the Higgs decay mode $pp \to H \to \gamgam$.
We will therefore primarily talk about photons, but we stress that our
results and conclusions apply equally to leptons.
Using a simplified calculation at NLO, we determine the scaling of power corrections
induced by the cuts. In particular, we discuss the dependence on the isolation method
and parameters, considering both fixed-cone and smooth-cone isolations.
We will find that the cuts induce power corrections that are parametrically enhanced,
and which can thus be significantly larger than for the case without cuts.
This enhancement is particularly severe for the case of $q_T$ subtractions with
smooth-cone isolation.
This has important ramifications for the numerical stability of the subtractions
in practical applications.
In fact, in \refscite{Grazzini:2015nwa, Grazzini:2017mhc}
it was already observed numerically that processes involving photon isolation
suffer from large enhanced power corrections, which is explained by our results.

Given the potentially significant size of the cut-induced power corrections,
it is essential to account for them. Since in general they are complicated and cut specific,
including them by an explicit analytic calculation (e.g.\ along the lines
of the inclusive ones discussed above) would be challenging and tedious.
Differential subtractions~\cite{Gaunt:2015pea} offer a way to avoid
the power corrections because they do not require the finite cutoff
that is necessary in the slicing approach.
Exploiting this, we propose a strategy to incorporate the measurement cuts exactly
such that the additional cut-induced power corrections are avoided.
It uses the Born-like measurement that appears in the singular
subtractions to separate the cut-induced power corrections from the
inclusive, cut-independent ones, where the former can be kept exactly while the
latter can be treated in the standard way.
We also show that in this way the projection-to-Born method~\cite{Cacciari:2015jma}
naturally appears as the special case where the inclusive, cut-independent power corrections
are fully known.

This paper is structured as follows.
In \sec{review}, we briefly review the $q_T$ and $\Tau_N$ subtraction formalism
and give an overview of different photon isolation methods.
We then provide a simple analytic study of the effect of both selection and isolation
cuts on the subtraction techniques in \sec{setup_NLO},
before verifying our results numerically in \sec{numerics}.
Finally in \sec{subtractions}, we discuss how to incorporate the additional
measurement cuts into the subtractions.
We conclude in \sec{conclusions}.

%%%%%%%%%%%%%%%%%%%%%%%%%%%%%%%%%%%%%%%%%%%%%%%%%%%%%%%%%%%%%%%%%%%%%%%%%%%%%%%%
\section{Review of subtractions and photon isolation}
\label{sec:review}
%%%%%%%%%%%%%%%%%%%%%%%%%%%%%%%%%%%%%%%%%%%%%%%%%%%%%%%%%%%%%%%%%%%%%%%%%%%%%%%%

%===============================================================================
\subsection[Review of \texorpdfstring{$q_T$ and $\Tau_N$}{qT and TauN} subtractions]
           {Review of $q_T$ and $\Tau_N$ subtractions}
\label{sec:review_subtractions}
%===============================================================================

In this section, we briefly review the $q_T$ and $\Tau_N$ subtraction methods.
For a detailed discussion we refer to \refcite{Gaunt:2015pea}.

We denote the relevant dimensionful resolution variable generically as $\Tau$
and its dimensionless version as $\tau$.
For the case of color-singlet production ($N = 0$), it can be chosen as
the total transverse momentum of the color-singlet final state,
$\Tau \equiv q_T^2$, which yields $q_T$ subtractions~\cite{Catani:2007vq}.
For $0$-jettiness subtractions, it is given by $0$-jettiness
(aka beam thrust) $\Tau \equiv \Tau_0$.
In terms of the hadronic final-state momenta $k_i$, these are defined as%
\footnote{For $0$-jettiness or beam thrust, one can define more generic
measures~\cite{Berger:2010xi, Stewart:2010tn, Jouttenus:2011wh}. We focus on $\Tau_0^{\rm lep}$,
whose power corrections are smaller than for other definitions \cite{Moult:2016fqy, Ebert:2018lzn}.}
\begin{alignat}{3} \label{eq:qT}
 & \Tau &&\equiv q_T^2 = \Bigl(\sum_i \vec k_{T,i} \Bigr)^2
 \,,\qquad &&
 \tau \equiv q_T^2/Q^2
%%%
\,,\\ \label{eq:Tau0}
%%%
 & \Tau &&\equiv \Tau_0^\lep
 = \sum_i \min \bigl\{ k_i^+ e^Y \,,\, k_i^- e^{-Y} \bigr\}
 \,,\qquad &&
 \tau \equiv \Tau_0^{\rm lep}/Q
\,.\end{alignat}
Here, the sums over real emissions $i$ in the final state. The
$k^+ = n \cdot k$ and $k^- = \bn \cdot k$ are lightcone momenta,
with $n^\mu = (1,0,0,1)$ and $\bn^\mu = (1,0,0,-1)$ being lightlike reference
vectors along the beam directions,
and $Q$ and $Y$ are the total invariant mass and rapidity of the Born (the color-singlet)
final state.

A key property of $\tau$ is that it is an IR-safe $N$-jet resolution variable,
i.e.\ it vanishes for the Born process and in the IR-singular limit where all real
emissions $k_i$ become soft or collinear.
We can thus write the cross section $\sigma(X)$
as an integral over the cross section differential in $\tau$,
%%%
\begin{equation}
\sigma(X)
= \int\!\df\tau\, \frac{\df\sigma(X)}{\df\tau}
= \sigma(X, \tau_\cut) + \int_{\tau_\cut}\!\df\tau\, \frac{\df\sigma(X)}{\df\tau}
\,,\end{equation}
%%%
where the cumulative cross section as a function of $\tau_\cut$ is defined as
%%%
\begin{equation}
\sigma(X, \tau_\cut)
= \int^{\tau_\cut} \!\df\tau\, \frac{\df\sigma(X)}{\df\tau}
\,.\end{equation}
%%%
Here, $X$ denotes all measurements. It includes the measurements performed on
the Born process, including any selection cuts on its constituents.
It also contains any additional cuts on the hadronic final state such as isolation cuts.

The slicing method is obtained by adding and subtracting a
global subtraction term $\sigma^\sub(X,\tau_\cut)$,
%%%
\begin{align} \label{eq:nsub_master}
\sigma(X)
&= \sigma^\sub(X, \tau_\cut)
+ \int_{\tau_\cut} \!\df\tau\, \frac{\df\sigma(X)}{\df\tau}
+ \Delta\sigma(X, \tau_\cut)
\,, \nn \\
\Delta\sigma(X, \tau_\cut)
&= \sigma(X, \tau_\cut) - \sigma^\sub(X, \tau_\cut)
\,.\end{align}
%%%
Since $\tau$ vanishes by construction in the Born limit, the integral in \eq{nsub_master}
necessarily involves at least one resolved real emission, and hence $\df\sigma(X)/\df\tau$
can be calculated from the corresponding Born+1-parton calculation at one lower order.
The key requirement on $\sigma^\sub(X, \tau_\cut)$ is that it must contain the leading
terms in the $\tau_\cut\to 0$ limit. If that is the case,
then $\Delta\sigma(X, \tau_\cut)$ is a power correction of $\ord{\tau_\cut}$ which vanishes as
$\tau_\cut\to 0$ and hence it can be neglected for sufficiently small $\tau_\cut$.

To construct $\sigma^\sub$ and study the size of $\Delta\sigma$, it is useful
to expand the differential cross section and its cumulant for $\tau \ll 1$
and correspondingly $\tau_\cut \ll 1$,
%%%
\begin{alignat}{2}\label{eq:xsec_expand}
\frac{\df \sigma(X)}{\df\tau}
&= \frac{\df\sigma^{(0)}(X)}{\df\tau} &&+ \sum_{m>0} \frac{\df\sigma^{(2m)}(X)}{\df\tau}
\,, \\ \nn
\sigma(X,\tau_\cut)
&= \sigma^{(0)}(X,\tau_\cut) &&+ \sum_{m>0} \sigma^{(2m)}(X,\tau_\cut)
\,,\end{alignat}
%%%
where the different contributions scale as
%%%
\begin{alignat}{2} \label{eq:xs_scaling}
 \frac{\df\sigma^{(0)}(X)}{\df\tau} &\sim \delta(\tau) + \sum_{j \geq 0} \biggl[\frac{\ln^j\tau}{\tau}\biggr]_+
\,, \qquad
\sigma^{(0)}(X, \tau_\cut) &&\sim  1 + \sum_{j \geq 0} \ln^{1+j}\tau_\cut
%%%
\,,\nn\\
%%%
 \tau \frac{\df\sigma^{(2m)}(X)}{\df\tau} &\sim \sum_{j\ge0} \tau^m \ln^j\tau
\,, \qquad\qquad\quad
 \sigma^{(2m)}(X,\tau_\cut) &&\sim \sum_{j\ge0} \tau_\cut^m \ln^j\tau_\cut
\,.\end{alignat}
%%%
The $\df\sigma^{(0)}/\df\tau$ and $\sigma^{(0)}(\tau_\cut)$
are the leading-power (LP) or singular terms, as they diverge as $1/\tau$ for $\tau \to 0$.
In particular, they fully capture the cancellation of virtual and real IR divergences,
which is encoded in the $\delta$ and plus distributions.
The $\df\sigma^{(2m)}/\df\tau$ with $m>0$ contain at most integrable divergences for $\tau\to0$,
and correspondingly $\sigma^{(2m)}(\tau_\cut\to 0) \to 0$.
They are thus referred to as nonsingular or power-suppressed corrections.

For \eq{nsub_master} to provide a viable subtraction, $\sigma^\sub(X, \tau_\cut)$
must at least contain the singular terms, i.e., we require
%%%
\begin{align}
 \sigma^\sub(X,\tau_\cut) &= \sigma^{(0)}(X,\tau_\cut) \bigl[ 1 + \cO(\tau_\cut) \bigr]
\,.\end{align}
%%%
The correction term in \eq{nsub_master} then scales as a power correction
%%%
\begin{align}
\Delta\sigma(X,\tau_\cut) &
 = \sigma(\tau_\cut) - \sigma^\sub(X,\tau_\cut)
 = \cO\bigl(\tau^{m}_\cut\bigr)
\,,\end{align}
%%%
where $m$ is determined by the first term in the sum in \eq{xsec_expand}
that is not contained in $\sigma^\sub$.

For inclusive Higgs and Drell-Yan production, the sum in \eq{xsec_expand}
starts with $m = 1$ for both $q_T$ \cite{Ebert:2018lzn}
and $\Tau_0$ \cite{Gaunt:2015pea, Moult:2016fqy, Boughezal:2016zws, Moult:2017jsg}.
In these cases, the full $\cO(\tau_\cut^1)$ correction is known at NLO
\cite{Boughezal:2018mvf,Ebert:2018gsn,Ebert:2018lzn} and can be included
in $\sigma^\sub$ such that $\Delta\sigma \sim \cO(\tau_\cut^2)$.
In \sec{setup_NLO}, we will determine the scaling of $\Delta\sigma$
in the presence of selection and isolation cuts.

%===============================================================================
\subsection{Review of photon isolation}
\label{sec:review_isolation}
%===============================================================================

Photon production at hadron colliders such as the LHC is dominated by \emph{secondary} photons
arising from the decay of hadrons inside final-state jets, in particular $\pi^0,\eta \to \gamgam$,
whereas one is interested in \emph{prompt} photons directly produced in hard interactions.
Experimentally, secondary photons can be efficiently suppressed using the shape
of the electromagnetic showers in the calorimeter, see e.g.\ \refcite{Aad:2019tso}.
This is supplemented by an additional cone isolation which restricts the transverse energy inside a fixed cone of radius $R$ around the photon,
%%%
\begin{align} \label{eq:cone_isolation}
 \sum_{i:\,d(i,\gamma) \le R} E_T^i \le \ETiso
\,.\end{align}
%%%
Here, the sum runs over all identified hadrons $i$ with momenta $k_i$, $E_T^i \equiv E_T(k_i)$ is their transverse energy, and the distance measure between two particles $i$ and $j$
is as usual given in terms of their difference in azimuth and pseudorapidity,
%%%
\begin{align}
 d(i,j) = \sqrt{(\phi_i-\phi_j)^2 + (\eta_i - \eta_j)^2}
\,.\end{align}
The isolation energy $\ETiso$ is typically chosen as either a fixed value
or relative to the photon transverse energy, $\ETiso = \eps\,p_{T\gamma}$.

Theory predictions employing this fixed-cone isolation require the use
of photon fragmentation functions $D_q$ to cancel collinear singularities
arising from collinear quark splittings $q \to q + \gamma$.
This is analogous to the absorption of collinear singularities
from initial-state splittings into parton distribution functions.
The fragmentation functions are nonperturbative objects and have been determined from data
\cite{Aurenche:1992yc,Gluck:1992zx,Bourhis:1997yu,GehrmannDeRidder:1997gf}.
After their inclusion, quark fragmentation factorizes into a nonperturbative
and perturbative piece, allowing for an infrared-safe calculation \cite{Catani:1998yh,Catani:2002ny}.

Currently, the fragmentation functions $D_q$ are only poorly constrained from data,
yielding large theory uncertainties.
Furthermore, for tight isolation cuts with small ${R\ll1}$ one encounters
large logarithms $\ln(R)$ which can render the perturbative calculation unstable \cite{Catani:2002ny}.
Their resummation has been addressed e.g. in \refscite{Catani:2013oma, Balsiger:2018ezi}.

To avoid the added complications of nonperturbative fragmentation functions,
perturbative calculations often employ
the smooth-cone isolation proposed by Frixione \cite{Frixione:1998jh},
as used e.g.\ in the NNLO calculations of direct diphoton production
in \refscite{Catani:2011qz, Campbell:2016yrh, Catani:2018krb}.%
\footnote{One can also employ a hybrid approach by combining smooth-cone isolation with radius $R_0$
with a fixed-cone isolation of larger radius $R \ll R_0$, as used e.g.\ in the NNLO calculation
of direct photon production in \refcite{Chen:2019zmr}.
}
Frixione isolation modifies \eq{cone_isolation} to
%%%
\begin{align} \label{eq:Frixione_isolation}
\sum_{i:\,d(i,\gamma) \le r} E_T^i \le \ETiso \, \chi(r) \qquad \forall r \le R
\,,\end{align}
%%%
where $\chi(r)$ is a function that vanishes as $\chi(r\to0)\to0$,
and $\ETiso$ can again be chosen as a fixed value or relative to the photon momentum,
$\ETiso = \eps\,p_{T\gamma}$.
This isolation constraint becomes stronger the closer the hadrons are to the photon.
In particular, it fully suppresses radiation exactly collinear to the photon,
and hence removes the collinear singularities from $q\to q+\gamma$ splittings.
On the other hand, soft radiation with $E_T\to0$ is not vetoed,
which is crucial to not spoil the cancellation of soft divergences.
Thus, calculations employing Frixione isolation are infrared safe without the inclusion of fragmentation functions.
Due to finite detector resolution, this isolation cannot be implemented experimentally,
but it has been shown to yield results compatible (within theory uncertainties) to
fixed-cone isolation for sufficiently tight isolations \cite{Andersen:2014efa,Badger:2016bpw,Catani:2018krb}.

A common choice of $\chi(r)$ is given by
%%%
\begin{align} \label{eq:chi1}
\chi(r) = \biggl[\frac{1 - \cos(r)}{1-\cos(R)}\biggr]^n
\,,\end{align}
%%%
with the parameter $n>0$, and we will use this implementation for our numerical results in \sec{numerics}.
For the analytic study in \sec{setup_NLO}, we will instead use
%%%
\begin{align} \label{eq:chi2}
 \chi(r) = \Bigl(\frac{r}{R}\Bigr)^{2n}
\,,\end{align}
%%%
which is a good approximation of \eq{chi1} for $r,R\ll1$.

For illustration purpose, we will also consider a harsh isolation criterion,
where one completely vetoes any radiation inside the isolation cone,
implemented by restricting the
total hadronic transverse energy in the isolation cones to vanish,
%%%
\begin{align} \label{eq:harsh_isolation}
\sum_{i:\,d(i,\gamma) \le R} E_T^i = 0
\,.\end{align}
%%%
While this criterion is of course infrared unsafe, as even soft radiation is vetoed,
it will be useful to illustrate how factorization-violating effects can potentially arise.

Finally we note that recently a new isolation technique based
on jet substructure techniques was proposed in \refcite{Hall:2018jub}.
Here, one uses soft drop to identify ``photon jets'' that do not contain notable substructure
and defines these as isolated photons.
In the case of a single emission with momentum $k$ and distance $r < R$ from the photon,
this technique amounts to requiring that
%%%
\begin{align} \label{eq:softdrop_isolation}
k_T < p_{T\gamma} \frac{z_{\rm cut} (r/R)^\beta}{1-z_{\rm cut} (r/R)^\beta}
\,,\end{align}
%%%
where $R$ is size of the isolation cone, and $z_{\rm cut} < 1/2$ and $\beta$ are soft-drop parameters.
As discussed in \refcite{Hall:2018jub}, \eq{softdrop_isolation} is equivalent
to the Frixione isolation in \eqs{Frixione_isolation}{chi2}
in the limit of small $z_{\rm cut}$ or $r/R$ if one identifies
$\ETiso = z_{\rm cut} \, p_{T\gamma}$ and $\beta = 2n$.
Hence we will not discuss this technique separately.

%%%%%%%%%%%%%%%%%%%%%%%%%%%%%%%%%%%%%%%%%%%%%%%%%%%%%%%%%%%%%%%%%%%%%%%%%%%%%%%%
\section{Effect of isolation and fiducial cuts on singular cross sections}
\label{sec:setup_NLO}
%%%%%%%%%%%%%%%%%%%%%%%%%%%%%%%%%%%%%%%%%%%%%%%%%%%%%%%%%%%%%%%%%%%%%%%%%%%%%%%%

In this section, we present analytic arguments to derive the size of power corrections
induced by kinematic selection and isolation cuts.
For simplicity we consider the case of color-singlet production, though our
conclusions on the parametric size of the cut-induced power corrections also
apply to the $N$-jet case.
The general setup to calculate such corrections is presented in \sec{setup},
where we largely follow the strategy in \refscite{Ebert:2018lzn, Ebert:2018gsn}.
Kinematic selection cuts are discussed in \sec{pc_acceptance}
and isolation cuts are discussed in \sec{pc_isolation}.
We will numerically verify our results in \sec{numerics}.

%===============================================================================
\subsection{General setup}
\label{sec:setup}
%===============================================================================

We consider the production of a generic color-singlet final state $L$
at fixed total invariant mass $Q$ and rapidity $Y$, and in the presence of
additional cuts $X$. In \sec{review_subtractions} we kept $Q$ and $Y$ as part of
$X$. For our discussion here it is important to explicitly separate
the measurements $Q$ and $Y$ that parametrize the Born phase space from the
additional cuts $X$. We also measure a 0-jet resolution variable
$\Tau$ that is only sensitive to additional radiation and thus vanishes at LO.
Later on, we will specify to $\Tau \equiv q_T^2$ and $\Tau \equiv \Tau_0$.
The Born process is denoted by
%%%
\begin{align} \label{eq:proc_LO}
 a(p_a) + b(p_b) \to L(\{p_i\})
\,,\end{align}
%%%
where $a$ and $b$ are the flavors of the incoming partons, which carry momenta $p_a$ and $p_b$,
the color-singlet final state is composed of particles with individual momenta $\{p_i\}$,
and we denote the total momentum of $L$ by $q^\mu = \sum_i p_i^\mu$.
The Born cross section is given by
%%%
\begin{align} \label{eq:sigmaLO}
\frac{\df\sigma^\LO(X)}{\df Q^2 \df Y \df\Tau} &
 = \frac{f_a(x_a) f_b(x_b)}{2 x_a x_b \Ecm^4} \Msquared^\LO(Q,Y; X) \,\delta\bigl(\Tau\bigr)
\,\end{align}
%%%
where $f_a$ and $f_b$ are the parton distribution functions for particles $a$ and $b$,
$\Ecm$ is the hadronic center-of-mass energy, and the LO partonic cross section
$\Msquared^\LO(Q,Y;X)$ is given by
%%%
\begin{align} \label{eq:A_LO}
\Msquared^\LO(Q,Y; X)
&= \int\!\df\Phi_L(p_a+p_b) \, \Abs{\cM^\LO_{ab\to L}(p_a, p_b; \{p_i\})}^2 f_X(\{p_i\})
%%%
\,,\\
%%%
 \label{eq:phiL}
  \df\Phi_L(q) &= \biggl[\prod_i \frac{\df^4 p_i}{(2\pi)^3}\, \delta_+(p_i^2 - m_i^2) \biggr]
\, (2\pi)^4 \delta^{(4)}\Bigl(q - \sum_i p_i\Bigr)
\,.\end{align}
In \eq{A_LO}, $f_X(\{p_i\})$ implements the cuts on the final state momenta $\{p_i\}$,
which are kept implicit in the phase-space integral $\df\Phi_L(q)$.
In \eq{phiL}, $\delta_+(p^2 - m^2) = \theta(p^0) \delta(p^2-m^2)$ are on-shell $\delta$ functions.
Finally, the incoming momenta of the Born process are given by
%%%
\begin{align}
p_a^\mu &= x_a \Ecm \frac{n^\mu}{2} = Q e^{+Y} \frac{n^\mu}{2}
\,,\qquad
p_b^\mu = x_b \Ecm \frac{\bn^\mu}{2} = Q e^{-Y} \frac{\bn^\mu}{2}
\,,\end{align}
%%%
where as before $n^\mu = (1,0,0,1)$ and $\bn^\mu = (1,0,0,-1)$
are lightlike reference vectors along the beam directions.

Next, we consider the correction to \eq{proc_LO} from a single real emission,
%%%
\begin{align} \label{eq:proc_NLO}
 a'(p'_a) + b'(p'_b) \to L(\{p_i'\}) + k(k)
\,,\end{align}
where $k^\mu$ is the momentum of the emitted parton.
The resulting cross section is given by
\begin{align} \label{eq:sigma1}
 \frac{\df\sigma^{\rm real}(X)}{\df Q^2 \df Y \df\Tau} &
 = \int\!\frac{\df^d k}{(2\pi)^d} (2\pi) \delta_+(k^2) \,
   \frac{f_{a'}(\zeta_a) f_{b'}(\zeta_b)}{2 \zeta_a \zeta_b \Ecm^4}
   \delta\bigl[\Tau - \hat\Tau(k)\bigr]
   \\*\nn&\quad\times
  \int\df\Phi_L(p'_a+p'_b-k) \, \Abs{\cM(p'_a, p'_b; k, \{p_i'\})}^2 \, f_X(k, \{p_i'\})
\,.\end{align}
Here, $\cM$ is the matrix element for the process in \eq{proc_NLO},
including the relevant strong coupling constant $\as$ and renormalization scale $\mu^{d-4}$,
and $\hat\Tau(k)$ is the measurement operator that determines the value of $\Tau$ as a function of $k$.
The measurement function $f_X$ now acts on both $k$ and $\{p_i'\}$.
The incoming momenta are now fully determined in terms of $k$ and the measurements
of $Q$ and $Y$ as
%%%
\begin{align} \label{eq:p_ab}
 {p'_a}^\mu &= \zeta_a \Ecm \frac{n^\mu}{2}
             = \Bigl(k^- +  e^{+Y} \sqrt{Q^2 + k_T^2}\Bigr) \frac{n^\mu}{2}
\,,\nn\\
 {p'_b}^\mu &= \zeta_b \Ecm \frac{\bn^\mu}{2}
             = \Bigl(k^+ +  e^{-Y} \sqrt{Q^2 + k_T^2} \Bigr) \frac{\bn^\mu}{2}
\,.\end{align}
%%%
The restriction that $\zeta_{a,b} \in [0,1]$ is kept implicit in the support of the PDFs.

Resolution variables $\Tau$ sensitive to soft emission, $k^\mu\to0$,
and collinear emissions, $n \cdot k \to 0$ or $\bn \cdot k \to 0$, become singular in these limits.
Following the strategy of \refscite{Ebert:2018lzn,Ebert:2018gsn}, we can use
the SCET power expansion to organize the expansion of the cross section in the $\Tau\to 0$ limit
by considering the relevant collinear and soft scalings of $k^\mu$.
Resolution variables insensitive to the transverse momentum $k_T$ are described by SCET$_{\rm I}$,
where the appropriate modes are
\begin{alignat}{3} \label{eq:modes_SCET1}
 &\text{$n$-collinear}:   \quad&& k_n   \sim Q \, (\lambda^2, 1, \lambda)  &&
 \quad\Rightarrow\quad n \cdot k \ll k_T \ll \bn \cdot k
\,,\\\nn
 &\text{$\bn$-collinear}: \quad&& k_\bn \sim Q \, (1, \lambda^2, \lambda) &&
 \quad\Rightarrow\quad \bn \cdot k \ll k_T \ll n \cdot k
\,,\\\nn
 &\text{ultrasoft}:       \quad&& k_{us}\sim Q \, (\lambda^2, \lambda^2, \lambda^2)  &&
 \quad\Rightarrow\quad n \cdot k \,\sim k_T \,\sim \bn \cdot k
\,.\end{alignat}
Here, we use the lightcone notation $k^\mu = (k^+,k^-,k_T) = (n\cdot k,\bn\cdot k,k_T)$,
and $\lambda$ is a power-counting parameter.
For example, for $0$-jettiness $\Tau_0$ we have $\lambda \sim \sqrt{\Tau_0/Q}$.

Resolution variables resolving the transverse momentum $k_T$ fall into the realm of SCET$_{\rm II}$
and are characterized by the following modes,
\begin{alignat}{3} \label{eq:modes_SCET2}
 &\text{$n$-collinear}:   \quad&& k_n   \sim Q \, (\lambda^2, 1, \lambda) &&
 \quad\Rightarrow\quad n \cdot k \ll k_T \ll \bn \cdot k
\,,\\\nn
 &\text{$\bn$-collinear}: \quad&& k_\bn \sim Q \, (1, \lambda^2, \lambda) &&
 \quad\Rightarrow\quad \bn \cdot k \ll k_T \ll n \cdot k
\,, \\\nn
 &\text{soft}:            \quad&& k_s   \sim Q \, (\lambda, \lambda, \lambda) &&
 \quad\Rightarrow\quad n \cdot k \,\sim k_T \,\sim \bn \cdot k
\,.\end{alignat}
For example, for $q_T$ we have $\lambda \sim q_T/Q$.
\Eqs{modes_SCET1}{modes_SCET2} only differ by the scaling of soft and ultrasoft modes,
which will not change the analytic calculations here, only the resulting scaling
of power corrections in $\lambda$. In contrast, it does have a significant impact on the singular limit
of the matrix element itself, and for SCET$_{\rm II}$ it requires the use of rapidity regulators,
see \refscite{Ebert:2018lzn,Ebert:2018gsn} for more details.

Inserting the appropriate scalings of \eq{modes_SCET1} or \eq{modes_SCET2} into \eq{sigma1},
we can systematically expand the cross section in $\lambda$,
\begin{align} \label{eq:sigma2}
 \frac{\df\sigma(X)}{\df Q^2 \df Y \df\Tau} &
 = \underbrace{\frac{\df\sigma^{(0)}(X)}{\df Q^2 \df Y \df\Tau}}_{\sim \lambda^{-2}}
 + \sum_{m>0} \underbrace{\frac{\df\sigma^{(2m)}(X)}{\df Q^2 \df Y \df\Tau}}_{\sim \lambda^{2m-2}}
\,.\end{align}
As briefly reviewed in \sec{review_subtractions}, $\sigma^{(0)}$ is referred to
as leading-power or singular limit and contains the cancellation of all IR divergences.

It is easy to see from \eqs{sigma1}{p_ab} that the total momentum of $L$
reduces to its Born value at leading power, i.e.\
\begin{align} \label{eq:q_LP}
 q = p'_a + p'_b - k
 = \begin{pmatrix}
    \sqrt{Q^2 + k_T^2} \cosh(Y) \\ \kt \\ \sqrt{Q^2 + k_T^2} \sinh(Y)
   \end{pmatrix}
 = p_a + p_b + \cO(k_T)
\,.\end{align}
Hence at leading power, the phase space $\df\Phi_L$ in \eq{sigma1}
reduces to the Born phase space. Note also that the light-cone
coordinates $q^\pm$ only receive relative corrections of
$\ord{k_T^2/Q^2} \sim \ord{\lambda^2}$.

For the cuts $X$ to be infrared safe they must be insensitive to collinear splittings or
soft emissions, and hence reduce to their Born result at leading power. For the measurement
function $f_X$ in \eq{sigma1}, this implies
\begin{align} \label{eq:fX_LP}
 \df\Phi_L(p_a'+p_b'-k)  \, f_X(k, \{p_i'\})
 = \df\Phi_L(p_a+p_b) \, f_X(\{p_i\}) \times \bigl[1 + \cO(\lambda^m)\bigr]
\,.\end{align}
Here, on the right hand side the total momentum $q$ is replaced by its
Born value, $q \to p_a + p_b$, and the individual momenta $\{p_i'\}$ are
correspondingly evaluated in the Born limit $\{p_i'\} \to \{p_i\}$.
\Eqs{q_LP}{fX_LP} are key ingredients in the derivation of the factorization theorem
that predicts the leading singular terms $\sigma^{(0)}$.
In particular, they imply that to all orders in $\as$, the singular cross section
is only sensitive to the Born kinematics of the final state $L$.
The corrections beyond the Born approximation crucially depend on the
precise definition of $X$, but are always suppressed by $\cO(\lambda^m$),
where $m>0$ encodes the fact that $X$ is infrared safe.
For $m = 0$, $X$ would modify the leading singular behavior in $\Tau$ and hence
break the factorization for $\Tau$ and lead to a divergent result for the cross section.

The $\sigma^{(2m)}$ with $m>0$ in \eq{sigma2} denote power corrections to
the singular cross section $\sigma^{(0)}$.
They can be systematically computed by expanding all ingredients in \eq{sigma1}
to higher order in $\lambda$. The expansion of PDFs and matrix elements
in this approach has already been carried out for Higgs and Drell-Yan production
in \refscite{Ebert:2018lzn, Ebert:2018gsn}, which found that these corrections scale as
$\lambda^0$, i.e.\ the sum in \eq{sigma2} starts indeed with $m=1$ as expected
on general grounds.

Here, we extend these works by calculating the power corrections in \eq{fX_LP}
arising from the color-singlet phase space and additional measurement cuts.
They can be calculated by considering the cross section
%%%
\begin{align} \label{eq:sigma_iso1}
\frac{\df\sigma^\cuts(X)}{\df Q^2 \df Y \df\Tau}
&= \int\!\frac{\df^4 k}{(2\pi)^3}\, \delta_+(k^2) \,
   \frac{f_{a'}(\zeta_a) f_{b'}(\zeta_b)}{2 \zeta_a \zeta_b \Ecm^4}
   \delta\bigl[\Tau - \hat\Tau(k)\bigr]
\nn \\ &\quad\times
  \int\Bigl[ \df\Phi_L(p'_a+p'_b-k) \, f_X(k, \{p_i'\})
   - \df\Phi_L(p_a+p_b) \, f_X(\{p_i\}) \Bigr]
\nn \\ &\quad\qquad\times
   \Abs{\cM(p'_a, p'_b; k, \{p_i'\})}^2
\,,\end{align}
%%%
and expanding it in the power-counting parameter $\lambda$.
The difference in square brackets is the
difference between the exact and LP limit on the left and right-hand sides of \eq{fX_LP}.
Since it vanishes for $k\to 0$, the $k$ integral is IR finite and can be evaluated
in $d = 4$ dimensions.

Since \eq{sigma_iso1} contains the process-dependent matrix elements,
it is not possible to give a general result for the cut-induced power corrections.
To obtain a generic analytic understanding of their size,
in the following we assume that the squared matrix element only depends
on the total momentum $q^\mu$ of $L$ but not the individual momenta $\{p_i'\}$, i.e.,
we assume that
%%%
\begin{equation} \label{eq:M2_approx}
\Abs{\cM(p'_a, p'_b; k, \{p_i'\})}^2 \equiv \Abs{\cM(p'_a, p'_b; k, q}^2
\,.\end{equation}
%%%
This holds for Higgs production,
where due to the isotropic decay all details of the decay are encapsulated in the branching ratio.
While this is a crude approximation for more complicated processes such as direct photon production,
it is completely sufficient to obtain a qualitative understanding of the cut effects,
since their power suppression is determined by the term in square brackets in \eq{sigma_iso1}.

Using \eq{M2_approx} allows us to pull out the matrix element, so \eq{sigma_iso1} becomes
%%%
\begin{align} \label{eq:sigma_iso2}
\frac{\df\sigma^\cuts(X)}{\df Q^2 \df Y \df\Tau}
&= \int\!\frac{\df^4 k}{(2\pi)^3} \delta_+(k^2)
   \frac{f_{a'}(\zeta_a) f_{b'}(\zeta_b)}{2 \zeta_a \zeta_b \Ecm^4}
   \Abs{\cM(p'_a, p'_b; k, q)}^2 \delta\bigl[\Tau - \hat\Tau(k)\bigr]
   \times \Delta\Phi_X(Q,Y,k)
\nn\\
\Delta\Phi_X(Q, Y, k)
&= \int\Bigl[\df\Phi_L(p_a' + p_b' - k) \, f_X\bigl(k, \{p_i'\}\bigr)
   - \df\Phi_L(p_a + p_b) \, f_X\bigl(\{p_i\}\bigr) \Bigr]
\,,\end{align}
%%%
where $\Delta\Phi_X(Q,Y,k)$ fully contains the effect of the recoil due to the emission $k$
on $\df\Phi_L$ as well as the cuts $X$. Recall that $p_{a,b}'$ and $p_{a,b}$ are
determined in terms of $Q$, $Y$, and $k$.

Using \eq{sigma_iso2}, it is straightforward to deduce the scaling
of the cut-induced power corrections by expanding $\Delta\Phi_X$ to
the first nonvanishing order in $\lambda$, while keeping the remaining
terms in \eq{sigma_iso2} in the singular limit.
If $\Delta\Phi_X$ scales as $\cO(\lambda^{2m})$, then the resulting
power correction scales as $\df\sigma^\cuts(X)/\df\Tau \sim \lambda^{-2+2m}$.
More explicitly, for the two cases we are interested in we have
%%%
\begin{align} \label{eq:lep_power_corrections}
 \frac{\df\sigma^\cuts(X)}{\df Q^2 \df Y \df q_T^2}
 \sim \frac{1}{q_T^2} \biggl(\frac{q_T^2}{Q^2}\biggr)^m
\,,\qquad
 \frac{\df\sigma^\cuts(X)}{\df Q^2 \df Y \df \Tau_0}
 \sim \frac{1}{\Tau_0} \biggl(\frac{\Tau_0}{Q}\biggr)^{m}
\,.\end{align}
%%%
This should be compared to the normal power corrections that arise from
expanding the matrix elements, etc. for which $m=1$.
If the kinematic cuts or isolation requirements yield a larger value, $m>1$,
then their effects are parametrically suppressed compared to the normal power corrections,
while for $m<1$ they are parametrically enhanced, and for $m=0$ they would violate the factorization,
as explained above.
In the remainder of this section, we will determine $m$
for kinematic selection cuts and various photon isolation techniques.

%===============================================================================
\subsection{Kinematic selection cuts}
\label{sec:pc_acceptance}
%===============================================================================

We begin by discussing the power corrections induced by kinematic selection cuts.
As an illustrative example, we consider a color-singlet final state $L$ composed of two massless
particles with momenta $p_1$ and $p_2$, and impose a minimum transverse momentum cut on both particles,
%%%
\begin{equation} \label{eq:pTcut}
 p_{T1} \,, p_{T2} \ge p_T^\min
\,.\end{equation}
%%%
This is the most common selection cut, which is practically always applied.
In addition, in practice one also requires cuts on the rapidities $y_{1,2}$,
which we neglect here for simplicity as they do not lead to qualitatively new features.

We write the total momentum $q$ and the individual momenta $p_{1,2}$ as
\begin{align} \label{eq:momenta}
 q^\mu &= \Bigl( \sqrt{Q^2 + q_T^2} \cosh(Y) \,,\, q_T ,\, 0 \,, \sqrt{Q^2 + q_T^2} \sinh(Y) \Bigr)
\,,\nn\\
 p_1^\mu &= p_T \bigl( \cosh(Y + \dY) \,,\, \cos\varphi \,,\, \sin\varphi \,,\, \sinh(Y + \dY) \bigr)
\,,\nn\\
 p_2^\mu &= q^\mu - p_1^\mu
\,.\end{align}
Here, $q^\mu$ is parameterized in terms of its invariant mass $Q$, rapidity $Y$,
and transverse momentum $q_T$, and using overall azimuthal symmetry
we choose to align the transverse momentum with the $x$ axis.
The massless momentum $p_1^\mu$ is expressed in terms of the angle $\varphi$
between its transverse momentum $\pt$ and $\qt$ and the rapidity difference $\dY = y_1 - Y$,
where $y_1$ is the rapidity of $p_1^\mu$.
Note that using this parameterization, for $q_T=0$ one has $y_{1,2} = Y \pm \dY$.
For simplicity of notation, we now identify $p_T \equiv p_{T1}$, while $p_{T2}$ is defined implicitly through \eq{momenta}.
Momentum conservation yields the relation
\begin{align} \label{eq:pT_gam}
 p_T = \frac{Q^2 / 2}{\sqrt{Q^2+q_T^2}\cosh(\dY) - q_T \cos\varphi}
\,.\end{align}
%%%
In terms of the above variables, the two-particle phase space in \eq{phiL} is given by
%%%
\begin{align} \label{eq:phi_gamgam}
 \df\Phi_L(q) &
 = \frac{\df^4 p_1}{(2\pi)^3} \delta_+(p_1^2) \,
   \frac{\df^4 p_2}{(2\pi)^3} \delta_+(p_2^2) \,
   (2\pi)^4 \delta^{(4)}(p_1 + p_2 - q)
 = \frac{p_T^2}{8\pi^2Q^2} \df\varphi \,\df\dY
\,.\end{align}
%%%
Integrating this differential phase space in the presence of the cut in \eq{pTcut} yields
%%%
\begin{align} \label{eq:phi_gamgam_pTmin_1}
  \Phi_L(q,p_T^\min) &
 = \int\df\Phi_L(q) \, \theta\bigl(p_{T1} - p_T^\min\bigr) \theta\bigl(p_{T2} - p_T^\min\bigr)
\\\nn&
 = 4 \int_0^\pi\df\varphi \int_0^\infty \df\dY \, \frac{p_T^2}{8 \pi^2 Q^2}
   \theta\bigl[\min\bigl\{p_T^2, p_T^2 - 2 p_T q_T \cos\varphi + q_T^2 \bigr\} - (p_T^\min)^2\bigr]
\,.\end{align}
%%%
In the second line, we combined the two cuts into one $\theta$ function,
and employed symmetry of the integrand.
Note that this integral is independent of the total rapidity $Y$.

\Eq{phi_gamgam_pTmin_1} depends on the total transverse momentum $q_T$ only
through the combinations $q_T^2$ and $q_T \cos\varphi$.
Naively, one may thus expect that in the expansion of $\Phi_L(q,p_T^\min)$,
all odd powers of $q_T$ vanish due to the integral over the azimuthal angle $\varphi$,
which would imply that the first power correction arises at $\cO(q_T^2)$.
However, the minimum in \eq{phi_gamgam_pTmin_1} explicitly breaks the azimuthal symmetry.
Concretely, in the limit $q_T \ll Q$, we have
\begin{align}
 \min\bigl\{p_T^2, p_T^2 - 2 p_T q_T \cos\varphi + q_T^2 \bigr\}
 = \begin{cases}
    p_T^2 \,, \hspace{3.15cm} \cos\varphi < 0 \\
    p_T^2 - 2 p_T q_T \cos\varphi \,, \qquad\cos\varphi \ge 0 \\
   \end{cases}
\,,\end{align}
up to corrections of $\ord{q_T^2/p_T^2}$, and it is clear that this result breaks the azimuthal symmetry, such that the $\varphi$ integral does not vanish.

Expanding \eq{phi_gamgam_pTmin_1} correspondingly in $q_T \ll p_T \sim Q$, we obtain the result
\begin{align} \label{eq:phi_gamgam_pTmin_2}
 \Phi_L(q,p_T^\min) &
 = \Phi^{(0)}_L(q,p_T^\min) + \Phi^{(1)}_L(q,p_T^\min) + \ord{q_T^2/Q^2}
\,,\end{align}
where the LP and NLP results are given by
\begin{align}
 \label{eq:phi_gamgam_pTmin_LP}
 \Phi^{(0)}_L(q,p_T^\min) & = \frac{\theta(Q -2 p_T^\min)}{8\pi} \sqrt{1 - (2 p_T^\min / Q)^2}
\,,\\\
 \label{eq:phi_gamgam_pTmin_NLP}
 \Phi^{(1)}_L(q,p_T^\min) &= -\frac{1}{2\pi^2} \frac{q_T}{Q} \frac{p_T^\min}{Q} \frac{\theta(Q -2 p_T^\min)}{\sqrt{1 - (2 p_T^\min / Q)^2}}
\,.\end{align}
These scale as $\cO[(q_T/Q)^0]$ and $\cO[(q_T/Q)^1]$, respectively.
These results can be easily verified by comparing against the numerical evaluation of the exact expression in \eq{phi_gamgam_pTmin_1}.

\begin{figure}[t]
 \centering
 \includegraphics[width=0.5\textwidth]{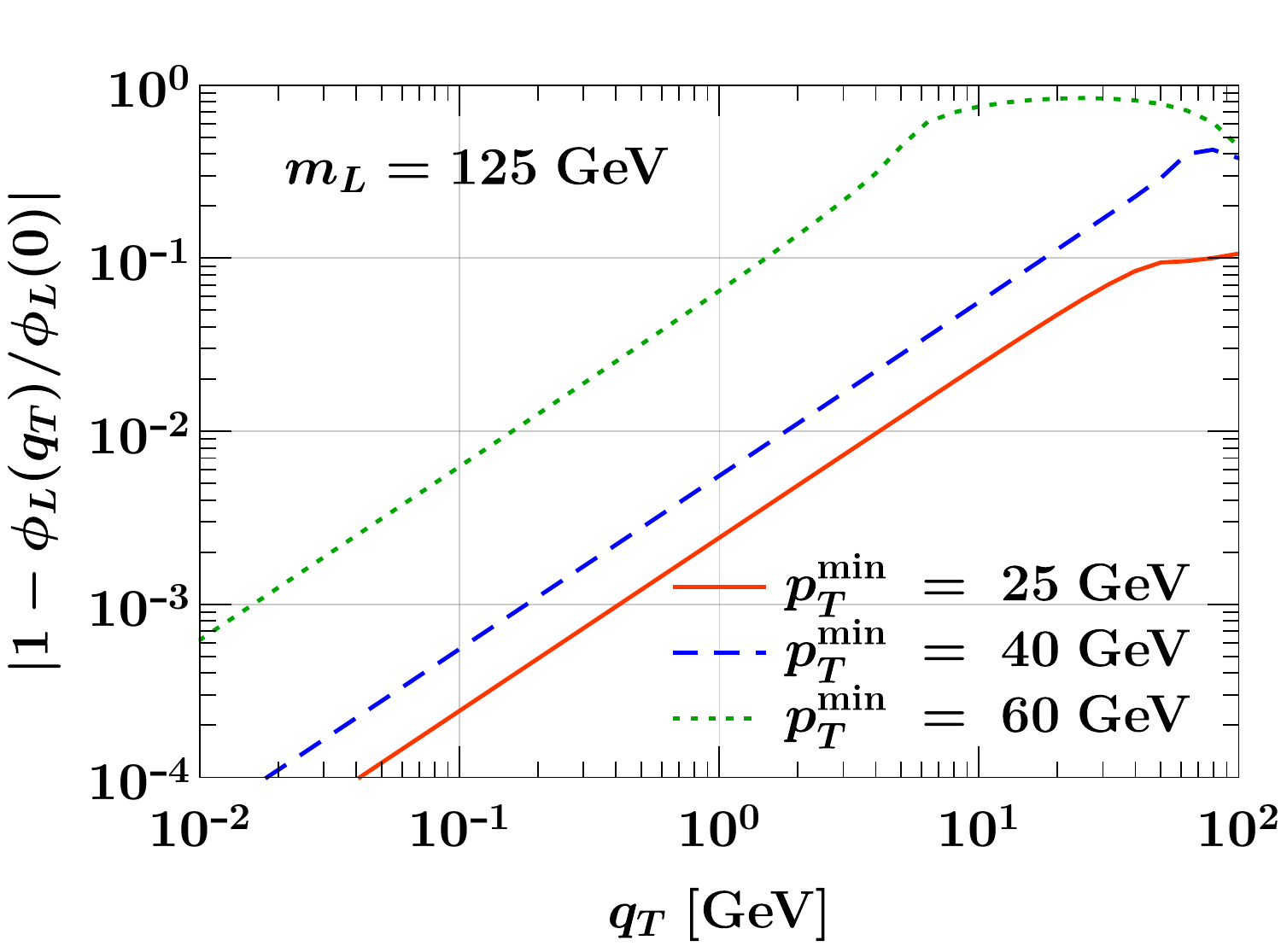}
 \caption{The two-particle phase space $\Phi_L(q_T)$ relative to its Born-level value $\Phi_L(0)$
 in the presence of a cut $p_T^\min$ on the individual momenta,
 as a function of the transverse momentum $q_T$.}
 \label{fig:phiGamGam_pTmin}
\end{figure}

For illustration, we show in \fig{phiGamGam_pTmin} the relative difference
between the exact phase space $\Phi_L$ and its Born approximation $\Phi_L^{(0)}$
in the presence of three different cuts $p_T^\min$, namely $p_T^\min = 25~\GeV$ (red solid),
$p_T^\min = 40~\GeV$ (blue dashed), and $p_T^\min = 60~\GeV$ (green dotted).
From the slope of each of the three curves, one can easily see the linear dependence on $q_T$,
and the slope is in perfect agreement with the result in \eq{phi_gamgam_pTmin_NLP}.

The function $\Delta\Phi_X$ that captures the power corrections induced by the $p_T^\min$ cut
is easily obtained by combining \eqs{sigma_iso2}{phi_gamgam_pTmin_2},
%%%
\begin{align} \label{eq:phi_gamgam_pTmin_3}
\Delta\Phi_{p_T^\min}
&= \Phi_L(q,p_T^\min) - \Phi^{(0)}_L(q,p_T^\min)
 = \cO\biggl(\frac{q_T}{Q} \frac{p_T^\min}{Q} \biggr)
\,.\end{align}
%%%
This linear dependence on $q_T$ translates into a relative power suppression
of $\cO(\lambda)$.
Thus the power corrections in \eq{lep_power_corrections} for a $p_T^\min$ cut
have $m = 1/2$ and scale as
%%%
\begin{align} \label{eq:lep_power_corrections_pTmin}
 \frac{\df\sigma^\cuts(X)}{\df Q^2 \df Y \df q_T^2}
 \sim \frac{1}{q_T^2} \frac{q_T}{Q}
\,,\qquad
 \frac{\df\sigma^\cuts(X)}{\df Q^2 \df Y \df \Tau_0}
 \sim \frac{1}{\Tau_0} \sqrt{\frac{\Tau_0}{Q}}
\,.\end{align}
%%%
Hence, compared to the normal case of $m = 1$, where the power corrections
scale as $q_T^2/Q^2$ and $\Tau_0/Q$, corresponding to $\ord{\lambda^2}$,
the power corrections induced by the kinematic selection
cuts are enhanced as $\cO(q_T/Q)$ and $\cO(\sqrt{\Tau_0/Q})$.
Intuitively, this arises from breaking
the azimuthal symmetry that is present in the Born process, but which is
explicitly broken by the recoil of the color-singlet system against the real emission.
Hence, additional kinematic selection cuts will generically induce enhanced power corrections
of $\ord{\lambda}$.

%===============================================================================
\FloatBarrier
\subsection{Photon isolation}
\label{sec:pc_isolation}
%===============================================================================

Next, we study the impact of photon isolation cuts on the power corrections.
To disentangle this effect from the fiducial cuts considered
in the previous section, we do not impose any other cuts besides the isolation.
We define an isolation function $f_\iso(k,p_\gamma)$ to evaluate to $1$
if the photon with momentum $p_\gamma$ is isolated from the emission with momentum $k$,
and to evaluate to $0$ otherwise.
The integrated phase space for diphoton production in the presence of such isolation,
as defined in \eq{sigma_iso2}, is given by
\begin{align} \label{eq:delta_iso1}
\Delta\Phi_\iso(Q,Y,k)
&= \int\Bigl[\df\Phi_L(p_a' + p_b' - k) \, f_\iso(k,p_1') f_\iso(k,p_2')
  - \df\Phi_L(p_a + p_b) \Bigr]
\,,\end{align}
where as before $p_{1,2}'$ are the momenta of the two photons,
$p_{a,b}'$ are the momenta of the incoming partons,
and $k$ is the momentum of the real emission.
To calculate the leading power behavior of \eq{delta_iso1},
it suffices to work in the singular limit of the phase space,
where the photons are back to back with total momentum $q^\mu = (Q \cosh Y, 0, 0, Q \sinh Y)$.
We parameterize their individual momenta by
\begin{align}
 p_1^\mu &= p_T \bigl(\cosh(Y+\dY)\,,\, +\cos(\varphi)\,,\, +\sin(\varphi)\,,\, \sinh(Y+\dY)\bigr)
\,,\nn\\
 p_2^\mu &= p_T \bigl(\cosh(Y-\dY)\,,\, -\cos(\varphi)\,,\, -\sin(\varphi)\,,\, \sinh(Y-\dY)\bigr)
\,,\end{align}
where the rapidity difference $\dY$ and the photon transverse momenta $p_T$ are related by
\begin{align}
 \cosh(\dY) = \frac{Q}{2 p_T}
\,.\end{align}
Using the expression \eq{phi_gamgam} for the diphoton phase space in the $q_T=0$ limit, we obtain
\begin{align} \label{eq:delta_iso2}
\Delta\Phi_\iso(Q,Y,k)
&= \frac{1}{32 \pi^2} \int \frac{\df \dY}{\cosh^2\dY} \int_{-\pi}^\pi\!\df\varphi\,
    \bigl[ f_\iso(k,p_1) f_\iso(k,p_2) -1 \bigr]
\,.\end{align}
The calculation can be further simplified by assuming that both photons are well separated,
such that their isolation cones never overlap with each other, and by assuming that the isolation
energies for both photons are identically chosen as $\ETiso$. Since we work in the Born limit here,
where $p_{T1} = p_{T2} \equiv p_T$, this assumption holds even if the isolation threshold is
chosen proportional to the photon momentum, $\ETiso = \eps p_{T}$.
This renders \eq{delta_iso2} symmetric in
both momenta $p_1$ and $p_2$, such that we obtain
\begin{align} \label{eq:delta_iso3}
\Delta\Phi_\iso(Q,Y,k)
&= \frac{1}{16 \pi^2} \int \frac{\df \dY}{\cosh^2\dY} \int_{-\pi}^\pi\!\df\varphi\,
    \bigl[f_\iso(k,p_1) - 1\bigr]
\,.\end{align}
In the following, we evaluate \eq{delta_iso3} for the different isolation techniques discussed in \sec{review_isolation}
to deduce the resulting power corrections.

%~~~~~~~~~~~~~~~~~~~~~~~~~~~~~~~~~~~~~~~~~~~~~~~~~~~~~~~~~~~~~~~~~~~~~~~~~~~~~~~
\subsubsection{Fixed-cone isolation}
\label{sec:cone_isolation}
%~~~~~~~~~~~~~~~~~~~~~~~~~~~~~~~~~~~~~~~~~~~~~~~~~~~~~~~~~~~~~~~~~~~~~~~~~~~~~~~

We first study the fixed-cone isolation as defined in \eq{cone_isolation}, for which we have
\begin{align} \label{eq:fiso_cone}
 f_{\rm cone}(k,p_\gamma) = 1 - \theta(k_T - \ETiso) \theta[R - d(k,p_\gamma)]
\,,\end{align}
such that the photon is considered isolated unless the parton is inside
the isolation cone of size $R$ and its transverse momentum exceeds the isolation energy $\ETiso$.
Evaluating \eq{delta_iso3} with \eq{fiso_cone} gives
\begin{align} \label{eq:delta_iso_exp1}
\Delta\Phi_{\rm cone}(Q,Y,k)
&= -\frac{\theta(k_T - \ETiso)}{16 \pi^2} \int \frac{\df \dY}{\cosh^2\dY}
    \int_{-\pi}^\pi \df\varphi \, \theta\Bigl[R - \sqrt{\varphi^2 + (Y + \dY - y_k)^2}\Bigr]
\nn\\&
 = -\frac{\theta(k_T - \ETiso)}{8 \pi^2} \int \frac{\df \dY}{\cosh^2\dY} \sqrt{R^2 - (Y + \dY - y_k)^2}
\,.\end{align}
Here, $y_k$ is the rapidity of $k$, and the range of the $\dY$ integral
is kept implicit from the support of the square root.
Note that \eq{delta_iso_exp1} is always negative,
because it arises from an additional phase space restriction.
For small $R^2\ll1$, it can be approximated by
\begin{align} \label{eq:delta_iso_exp2}
\Delta\Phi_{\rm cone}(Q,Y,k)
&= - \frac{R^2}{16\pi} \frac{\theta(k_T - \ETiso)}{\cosh^2(Y - y_k)} \times \bigl[ 1 + \cO(R^2) \bigr]
\,.\end{align}
This correction vanishes as $R\to0$, as in this limit the isolation turns off.

The nontrivial kinematic dependence of \eq{delta_iso_exp2} is entirely given by the denominator.
To understand the induced power corrections, we first rewrite it as
\begin{align} \label{eq:cosh}
 \frac{1}{\cosh^2(Y - y_k)}
 = \biggl( \frac{2 e^{Y-y_k}}{1 + e^{2(Y-y_k)}} \biggr)^2
 = \frac{4 k^+ k^-}{(k^- e^{-Y} + k^+ e^Y)^2}
\,.\end{align}
Using the power counting from \eqs{modes_SCET1}{modes_SCET2}, we find in the
$n$-collinear and soft limits the corrections
\begin{alignat}{3} \label{eq:pc_cosh}
 &\text{$n$-collinear}:   \quad&& k_n   \sim Q \, (\lambda^2, 1, \lambda) \,, &&
 \quad\Rightarrow\quad \frac{1}{\cosh^2(Y - y_k)} \quad\sim\quad \cO(\lambda^2)
 \,,\nn\\
 &\text{soft}:       \quad&& k_{s}\sim Q \, (\lambda, \lambda, \lambda) \,. &&
 \quad\Rightarrow\quad \frac{1}{\cosh^2(Y - y_k)} \quad\sim\quad \cO(\lambda^0)
\,,\end{alignat}
and the corresponding $\bn$-collinear and ultrasoft behavior follows trivially.
\Eq{pc_cosh} implies that only the (ultra)soft limit of \eq{delta_iso_exp2} can yield power corrections
that are enhanced relative to the normal $\cO(\lambda^2)$ corrections intrinsic to the factorization,
as the collinear corrections are always suppressed by (at least) $\cO(\lambda^2)$ as well.

From \eqs{delta_iso_exp2}{pc_cosh}, it follows immediately that the power
correction to the $q_T$ factorization from fixed-cone isolation is given by
\begin{align} \label{eq:pc_cone_qT}
 \frac{\df\sigma^{\rm (cone)}(X)}{\df Q^2 \df Y \df q_T^2} \sim \frac{R^2}{q_T^2} \theta(q_T - \ETiso)
\,.\end{align}
Thus, while the scaling behavior is that of a leading-power term, $1/q_T^2 \sim \lambda^{-2}$,
this correction only contributes to $q_T \ge \ETiso$, and hence is suppressed for sufficiently
large isolation energies. For a tight isolation, the effect can however become sizable.

The impact on the $\Tau_0$ subtraction is more involved, as it remains to integrate over $k$ against the $\Tau_0$ measurement.
To do so, we first note that the effect of collinear emissions
is always suppressed at least as $\Tau_0$ by virtue of \eq{pc_cosh}.
Thus, an enhanced power correction can only result from the soft limit,
which can be deduced by an explicit one-loop calculation.
The bare expression for the soft limit without isolation effect
is given by \cite{Ebert:2018lzn}
%%%
\begin{align} \label{eq:soft_NLO_1}
 \frac{\df\sigma^{\rm soft}}{\df Q^2 \df Y \df\Tau} &
 = \frac{\df \sigma^\LO}{\df Q^2 \df Y}
   \frac{\as \mathbf{C}}{\pi} \frac{e^{\eps \gamma_E} \mu^{2\eps}}{\Gamma(1-\eps)}
   \int_0^\infty\!\! \frac{\df k^+ \df k^-}{(k^+ k^-)^{1+\eps}}
\\\nn&\quad\times
   \, \Bigl[ \theta(e^{-Y} k^- - e^Y k^+) \delta(\Tau_0 - e^{Y} k^+)
   + \theta(e^Y k^+ - e^{-Y} k^-) \delta(\Tau_0 - e^{-Y} k^-) \Bigr]
\,,\end{align}
%%%
where $\mathbf{C} = C_F,C_A$ is the appropriate Casimir for quark annihilation and gluon fusion.
\Eq{soft_NLO_1} is the leading-power limit of the first line in \eq{sigma_iso2}
without taking effects from $\Delta\Phi_X$ into account.
By inserting \eq{delta_iso_exp2} into the integral in \eq{soft_NLO_1},
we can thus calculate the leading correction from the isolation.
Letting $\eps\to0$ and rescaling $k^\pm \to e^{\mp Y} k^\pm$ to remove any dependence on $Y$, we obtain
\begin{align} \label{eq:soft_NLO_2}
 \frac{\df\sigma^{\rm (cone)}}{\df Q^2 \df Y \df\Tau} &
 = - \frac{\df \sigma^\LO}{\df Q^2 \df Y} \times
   \frac{\as \mathbf{C}}{2\pi} \frac{R^2}{\pi}
   \int_0^\infty \frac{ \df k^-}{(\Tau_0 + k^-)^2} \theta[\Tau_0 k^- - (\ETiso)^2] \, \theta(k^- - \Tau_0)
\nn\\&
 = - \frac{\df \sigma^\LO}{\df Q^2 \df Y} \times
   \frac{\as \mathbf{C}}{2\pi} \frac{R^2}{\pi} \biggl[ \frac{2}{\Tau_0} \theta(\Tau_0-\ETiso) + \frac{\Tau_0}{\Tau_0^2 + (\ETiso)^2} \theta(\ETiso-\Tau_0) \biggr]
\,.\end{align}
In summary, the correction from fixed-cone isolation for $\Tau_0$ is given by
\begin{align} \label{eq:pc_cone_Tau0}
 \frac{\df\sigma^{\rm (cone)}(X)}{\df Q^2 \df Y \df \Tau_0} \sim
 \begin{cases} \displaystyle
  \frac{R^2}{\Tau_0} \biggl(\frac{\Tau_0}{\ETiso}\biggr)^2 \qquad,\quad \Tau_0 \le \ETiso
 \,,\\ \displaystyle
  \frac{R^2}{\Tau_0} \hspace{2.25cm},\quad \Tau_0 > \ETiso
 \,.\end{cases}
\end{align}
For $\Tau_0 > \ETiso$, this yields the leading-power $1/\Tau_0$ behavior,
albeit suppressed by $R^2$, while for $\Tau_0 < \ETiso$ this contribution
is highly suppressed as $(\Tau_0/\ETiso)^2$.

%~~~~~~~~~~~~~~~~~~~~~~~~~~~~~~~~~~~~~~~~~~~~~~~~~~~~~~~~~~~~~~~~~~~~~~~~~~~~~~~
\subsubsection{Smooth-cone isolation}
\label{sec:Frixione}
%~~~~~~~~~~~~~~~~~~~~~~~~~~~~~~~~~~~~~~~~~~~~~~~~~~~~~~~~~~~~~~~~~~~~~~~~~~~~~~~

Next, we consider the smooth-cone isolation, \eq{Frixione_isolation},
using the definition of \eq{chi2} for $\chi(r)$.
In this case, we have
\begin{align} \label{eq:fiso_Frix}
f_{\rm smooth}(k,p_\gamma)
&= 1 - \theta\bigl[k_T - \ETiso (d(k,p_\gamma) / R)^{2n} \bigr] \theta[R - d(k,p_\gamma)]
\nn\\&
 = 1 - \theta\bigl[d_\min - d(k,p_\gamma) \bigr]
\,,\end{align}
where
\begin{align} \label{eq:dmin}
 d_\min^2 = {\rm min}\bigl\{ R^2, R^2 (k_T / \ETiso)^{1/n} \bigr\}
\,.\end{align}
According to \eq{fiso_Frix}, the photon is considered isolated
unless the parton is inside the radiation cone and
its transverse energy exceeds the threshold value,
which itself depends on the distance between photon and parton.
\Eq{delta_iso3} in the presence of the isolation function \eq{fiso_Frix} can be evaluated similar to \eq{delta_iso_exp1} and yields
\begin{align} \label{eq:delta_iso_Frix1}
\Delta\Phi_{\rm smooth}(Q,Y, k)
&= -\frac{1}{16\pi^2} \int \frac{\df\dY}{\cosh^2(\dY)}
 \int_{-\pi}^{\pi} \df\varphi \,
 \theta\bigl[ d_\min - d(k,\gamma) \bigr]
\nn\\&
 = - \frac{R^2}{16\pi} \frac{(k_T/\ETiso)^{1/n}}{\cosh^2(Y - y_k)}
     \times \bigl[ 1 + \cO(d_\min^2) \bigr]
\,,\end{align}
where we expanded in small $d_\min$ and used that in the singular limit $k_T \ll Q$
the minimum in \eq{dmin} is always dominated by the second value.

From \eqs{delta_iso_Frix1}{pc_cosh}, it follows immediately that the power
correction to the $q_T$ factorization from smooth-cone isolation is given by
\begin{align} \label{eq:pc_smooth_qT}
 \frac{\df\sigma^{\rm (smooth)}(X)}{\df Q^2 \df Y \df q_T^2} \sim
 \frac{R^2}{q_T^2} \Bigl(\frac{q_T}{Q}\Bigr)^{1/n} \biggl(\frac{Q}{\ETiso}\biggr)^{1/n}
\,.\end{align}
Here, the overall $1/q_T^2$ arises from multiplying the leading-power singular with the isolation correction.
This result should be compared to the inclusive power corrections, which scale as $q_T^2 / Q^2$.
Hence, while the absolute size of the isolation effect is suppressed by $R^2$, it is enhanced
because the isolation energy $\ETiso$ is typically much smaller than the hard scale $Q$.
For $n>1/2$ the scaling in $q_T$ is also parametrically enhanced compared to the inclusive case,
and thus in practice the smooth-cone isolation can give sizable power corrections.

For $\Tau_0$, we have to distinguish that for collinear modes $k_T \sim \lambda Q \sim \sqrt{\Tau_0 Q}$,
while for ultrasoft modes $k_T \sim \lambda^2 Q \sim \Tau_0$. Taking \eq{pc_cosh} into account,
we can deduce the dominant corrections depending on the isolation parameter $n$ from \eq{delta_iso_Frix1} as
\begin{align} \label{eq:pc_smooth_Tau0}
 \frac{\df\sigma^{\rm (smooth)}(X)}{\df Q^2 \df Y \df \Tau_0} \sim
 \begin{cases} \displaystyle
  \frac{R^2}{\Tau_0} \, \Bigl(\frac{\Tau_0}{Q}\Bigr)^{1 + 1/(2n)} \Bigl(\frac{Q}{\ETiso}\Bigr)^{1/n} \qquad,\quad n < 1/2
  \,,\vspace{1ex} \\ \displaystyle
  \frac{R^2}{\Tau_0} \, \Bigl(\frac{\Tau_0}{Q}\Bigr)^{1/n} \Bigl(\frac{Q}{\ETiso}\Bigr)^{1/n} \hspace{1.55cm},\quad n > 1/2
 \,.\end{cases}
\end{align}
As for the $q_T$ case, there is an enhancement in $Q/\ETiso$
due to the typically small value for the isolation energy.
Furthermore, compared to the inclusive case where the correction scales as $\cO(\Tau^1)$,
the scaling in $\Tau_0$ is parametrically enhanced for $n>1$.
Hence, the relative parametric enhancement compared to the normal case
turns out to be more severe for $q_T$ than $\Tau_0$.

The results in \eqs{pc_smooth_qT}{pc_smooth_Tau0}
hold for $q_T < \ETiso$ or $\Tau_0 < \ETiso$, in which case the
minimum in \eq{dmin} is given by the second term,
which then induces the $k_T$ dependence of \eq{delta_iso_Frix1}.
For the opposite case of $q_T > \ETiso$ or $\Tau_0 > \ETiso$,
the minimum in \eq{dmin} is instead given by $d_\min = R$,
such that smooth-cone isolation reduces to fixed-cone isolation.
Thus, we find that for $q_T > \ETiso$ or $\Tau_0 > \ETiso$, smooth-cone isolation
yields the same leading-power $1/q_T$ or $1/\Tau_0$ behavior as for fixed-cone isolation.

%~~~~~~~~~~~~~~~~~~~~~~~~~~~~~~~~~~~~~~~~~~~~~~~~~~~~~~~~~~~~~~~~~~~~~~~~~~~~~~~
\subsubsection{Harsh isolation.}
\label{sec:harsh_isolation}
%~~~~~~~~~~~~~~~~~~~~~~~~~~~~~~~~~~~~~~~~~~~~~~~~~~~~~~~~~~~~~~~~~~~~~~~~~~~~~~~

Finally, we consider the harsh isolation defined in \eq{harsh_isolation}, where
\begin{align}
 f_{\rm harsh}(k,p_\gamma) = 1 - \theta\bigl[R - d(k,p_\gamma)\bigr]
\,,\end{align}
which vetoes any radiation inside the isolation cone.
The corresponding result for \eq{delta_iso3} is easily obtained from \eq{delta_iso_exp2} by setting $\ETiso=0$,
\begin{align} \label{eq:delta_iso_harsh1}
\Delta\Phi_{\rm harsh}(Q,Y,k)
&= - \frac{R^2}{16\pi} \frac{\theta(k_T)}{\cosh^2(Y - y_k)} \times \bigl[ 1 + \cO(R^2) \bigr]
\,.\end{align}
%%%
The induced correction then follows directly from \eqs{pc_cone_qT}{pc_cone_Tau0} as
\begin{align}\label{eq:pc_harsh}
 \frac{\df\sigma^{\rm (harsh)}(X)}{\df Q^2 \df Y \df q_T^2} \sim \frac{R^2}{q_T^2}
\,,\qquad
 \frac{\df\sigma^{\rm (harsh)}(X)}{\df Q^2 \df Y \df \Tau_0} \sim \frac{R^2}{\Tau_0}
\,.\end{align}
This is a leading power (singular) effect, as the harsh isolation completely removes
part of the real emission phase space, namely the vicinity of the two photons,
and thus immediately breaks both factorization theorems, which rely on an
analytic integration over the full emission phase space.

%===============================================================================
\subsection{Factorization violation in photon isolation}
\label{sec:factorization_violation}
%===============================================================================

In this section, we briefly discuss a
potential source for factorization violation for isolation methods
when not carefully applying the isolation procedure.
In general, one only keeps events that satisfy
the chosen isolation criterion. The remaining events can then
still contain jets, as defined by a suitable jet algorithm applied after the isolation,
that are inside or overlapping the isolation cones, e.g.\ if the jets are sufficiently
soft.

Since any jet inside the isolation cone will typically be quite soft,
as part of the overall isolation procedure one can in principle also remove any jets
inside the isolation cone from further consideration, i.e., the events are kept
but the jets are not further considered for the calculation of physical quantities,
e.g.\ jet selection cuts. This approach is for example proposed in the original definition of
smooth-cone isolation in \refcite{Frixione:1998jh}.

For the purpose of the subtractions, it is however crucial to keep \emph{all}
reconstructed jets, or more generally all emissions, for the determination of the
resolution variable $\Tau$.
More generally, this applies to employing any factorization theorem,
irrespective of whether it is used for subtractions or resummation of large logarithms.
For example, recall the definition for $0$-jettiness, see \eq{Tau0}
\begin{align} \label{eq:Tau0_2}
 \Tau_0 = \sum_i \min \bigl\{ k_i^+ e^Y \,,\, k_i^- e^{-Y} \bigr\}
\,.\end{align}
Here, the sum $i$ runs over all particles $i$ in the final state, only excluding
the color-singlet final state,
which is critical for the derivation of the $\Tau_0$ factorization theorem.
Excluding any emissions inside the isolation cones from the sum in \eq{Tau0_2}
would thus change the definition of $\Tau_0$ and immediately violate
the $\Tau_0$ factorization theorem.
For example, at one loop, where one has only one real emission, excluding jets
inside the isolation cones is equivalent to excluding the emission.
As far as calculating $\Tau_0$ is concerned, this exactly corresponds to the harsh isolation
defined in \eq{harsh_isolation}. As discussed in \sec{harsh_isolation},
this induces leading-power corrections, which exactly corresponds
to breaking the factorization.

For $q_T$ subtraction, one can trivially avoid this problem by determining
$q_T$ directly from the color-singlet final state $L$, i.e.\ $q_T \equiv q_{T,L}$.
On the other hand, if $q_T$ is obtained from the sum of all real emissions,
$q_T = |\sum_i \vec k_{T,i}|$, then as for $\Tau_0$, the sum over $i$ must not exclude
emissions inside the isolation cones to not violate the factorization.

Lastly, we point out that this leads to a trivial yet dangerous pitfall in the
calculation of power corrections.
For example, to calculate the NLO cross section for $pp\to H$ using $\Tau_0$ subtractions,
one would use $pp\to H+j$ at LO to calculate the power corrections or the
above-cut contributions in the slicing approach.
Naively applying the smooth-cone isolation including the discussed treatment of jets to
the resulting $H+j$ events, one would classify
all events where the emitted parton falls inside the isolation cone as $0$-jet events,
which depending on the used tool might be discarded in a $pp\to H+j$ calculation,
where at least one jet is required at Born level.
We have explicitly checked that this is the case for MCFM8
\cite{Campbell:1999ah, Campbell:2010ff, Campbell:2015qma, Boughezal:2016wmq}.
To not violate the subtraction method, it is however mandatory to keep all such events,
and we have turned off this mechanism in MCFM8 to obtain the correct results
for our numerical studies in \sec{numerics}.
(This does not impact the NLO calculations in MCFM8 itself, which keeps the
$H+j$ events that are otherwise classified as $0$-jet events.)

%%%%%%%%%%%%%%%%%%%%%%%%%%%%%%%%%%%%%%%%%%%%%%%%%%%%%%%%%%%%%%%%%%%%%%%%%%%%%%%%
\FloatBarrier
\section{Numerical results}
\label{sec:numerics}
%%%%%%%%%%%%%%%%%%%%%%%%%%%%%%%%%%%%%%%%%%%%%%%%%%%%%%%%%%%%%%%%%%%%%%%%%%%%%%%%

To validate our findings and assess the importance of the discussed power corrections,
we numerically study the $q_T$ and $\Tau_0$ spectrum at NLO$_0$,%
\footnote{We use this nomenclature to stress that this is part of the NLO correction
to the $0$-jet Born process $pp \to L$, rather than considering it as the LO$_1$ result
for the Born+$1$-parton process $pp \to L+j$.}
for direct diphoton production, $pp\to\gamgam$,
and for gluon-fusion Higgs production in the diphoton decay mode, $pp \to H \to \gamgam$,
using different photon acceptance cuts and isolation methods.
In all cases, we compare the full QCD result obtained from MCFM8
\cite{Campbell:1999ah, Campbell:2010ff, Campbell:2015qma, Boughezal:2016wmq}
against the predicted singular spectrum obtained from SCETlib \cite{SCETlib}.
For both processes, we use the PDF4LHC15\_nnlo\_mc \cite{Butterworth:2015oua} PDF set
and fix the factorization and renormalization scales to $\mu_f = \mu_r = m_H = 125~\GeV$.

To present our results, we normalize the cross section with the cuts $X$
to the LO cross section $\sigma^\LO(X^\LO)$ and split it into singular
and nonsingular contributions,
%%%
\begin{align} \label{eq:sigma_plot}
 \frac{\df \hat\sigma^{\rm full}(X)}{\df \Tau}
 \equiv \frac{1}{\sigma^\LO(X^\LO)} \frac{\df \sigma(X)}{\df \Tau} &
 = \frac{\df\hat\sigma^{\rm sing}}{\df \Tau} + \frac{\df\hat\sigma^{\rm nons}(X)}{\df \Tau}
 \nn\\&
 = \frac{\df\hat\sigma^{\rm sing}}{\df \Tau} + \frac{\df\hat\sigma^{\rm nons}}{\df \Tau}
   + \frac{\df\Delta\hat\sigma^{\rm nons}(X)}{\df \Tau}
\,.\end{align}
Here, $X^\LO$ indicates that the cuts only act on the Born kinematics of the
produced diphoton system, which in particular implies that there are no isolation effects.
For the normalized singular cross section
$\hat\sigma^{\rm sing} = \sigma^{(0)}(X^\LO) / \sigma^\LO(X^\LO)$,
the dependence on $X^\LO$ fully cancels since the LP cross section only depends on
the Born-level cuts $X^\LO$.
The nonsingular cross section $\hat\sigma^{\rm nons}(X) = \hat\sigma^{\rm full}(X) - \hat\sigma^{\rm sing}$
contains all power-suppressed contributions. In the second line, we have further split this piece
into the power corrections $\df\hat\sigma^{\rm nons}$ that are already present without additional cuts%
\footnote{When considering the effect of isolation cuts,
this piece corresponds to the nonsingular contribution without isolation but with
a potential $p_T^\min$ cut.}
and the additional power corrections $\df\Delta\hat\sigma^{\rm nons}(X)$ that are
induced by the cuts $X$. Comparing these two thus gives a direct indication of
their relative importance.

For Higgs production, we work in the on-shell limit where the invariant mass is fixed to $Q = m_H = 125~\GeV$,
while for diphoton production we restrict $Q = m_\gamgam = 120{-}130~\GeV$ such that $m_\gamgam \sim m_H$.
In both cases, we are inclusive over the rapidity $Y$ of the final state.
For direct photon production, we furthermore restrict ourselves to the $q \bar q \to \gamgam + g$ channel
to avoid contributions from the fragmentation process $q g \to \gamma + q (\to q + \gamma)$.
This allows us to obtain results without any photon isolation or fragmentation functions,
and thus compare the results with and without photon isolation.
Since direct diphoton production is divergent in the forward limit $p_T \to 0$,
we always impose selection cuts $p_T > p_T^\min = 25~\GeV$ to obtain a finite cross section.
This is not necessary for Higgs production, which we can also consider without any
photon selection cuts.

%===============================================================================
\subsection{Kinematic selection cuts}
\label{sec:numerics_pc_acceptance}
%===============================================================================

We first study the effect of fiducial cuts by comparing $pp \to H \to \gamgam$
with a lower cut on the photon transverse momenta, $p_T > p_T^\min = 25~\GeV$,
to the inclusive case without such a cut. As mentioned above, the same comparison
cannot be performed for direct diphoton production, since it diverges in the forward limit.

\begin{figure}[t]
 \centering
 \includegraphics[width=0.48\textwidth]{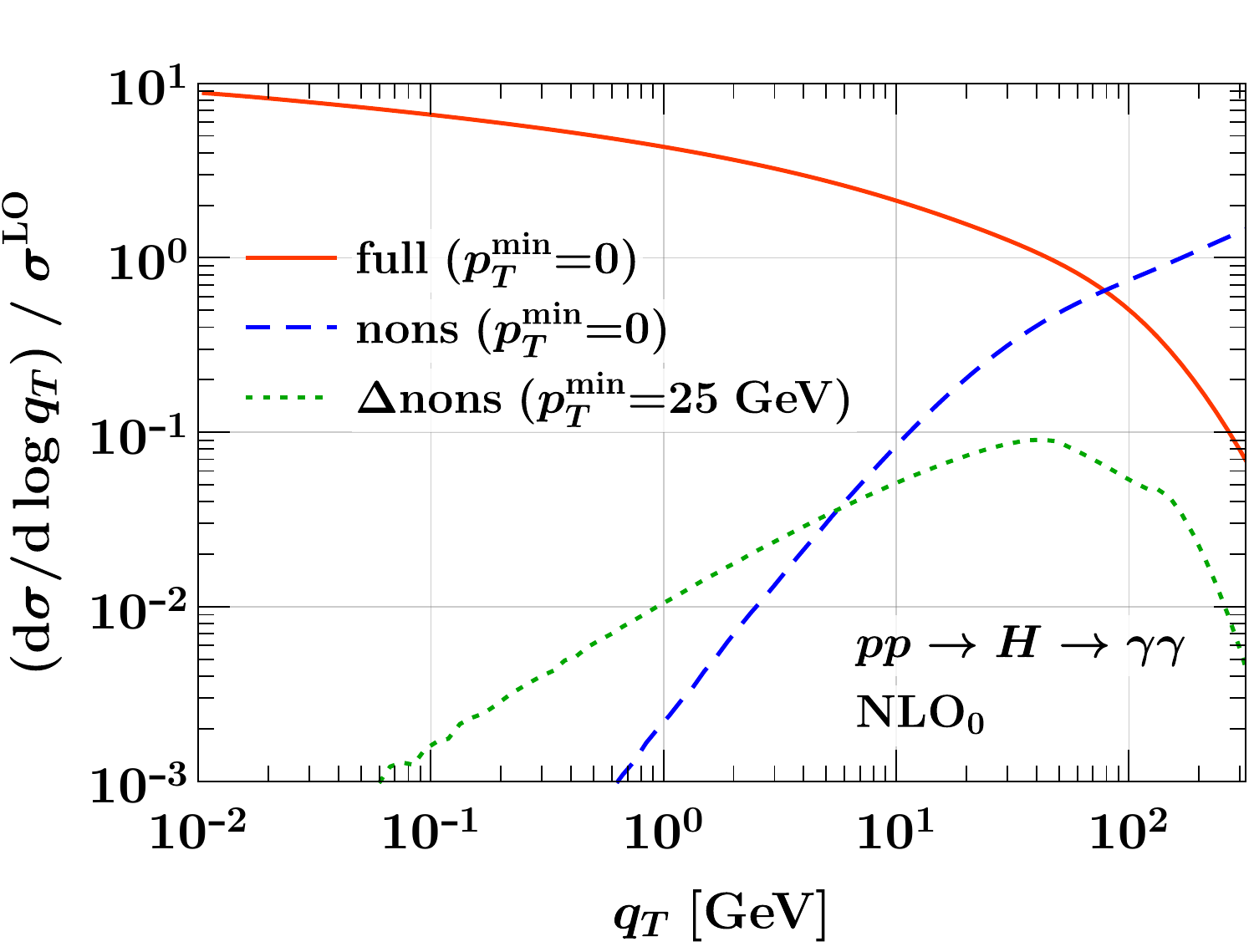}
 \quad
 \includegraphics[width=0.48\textwidth]{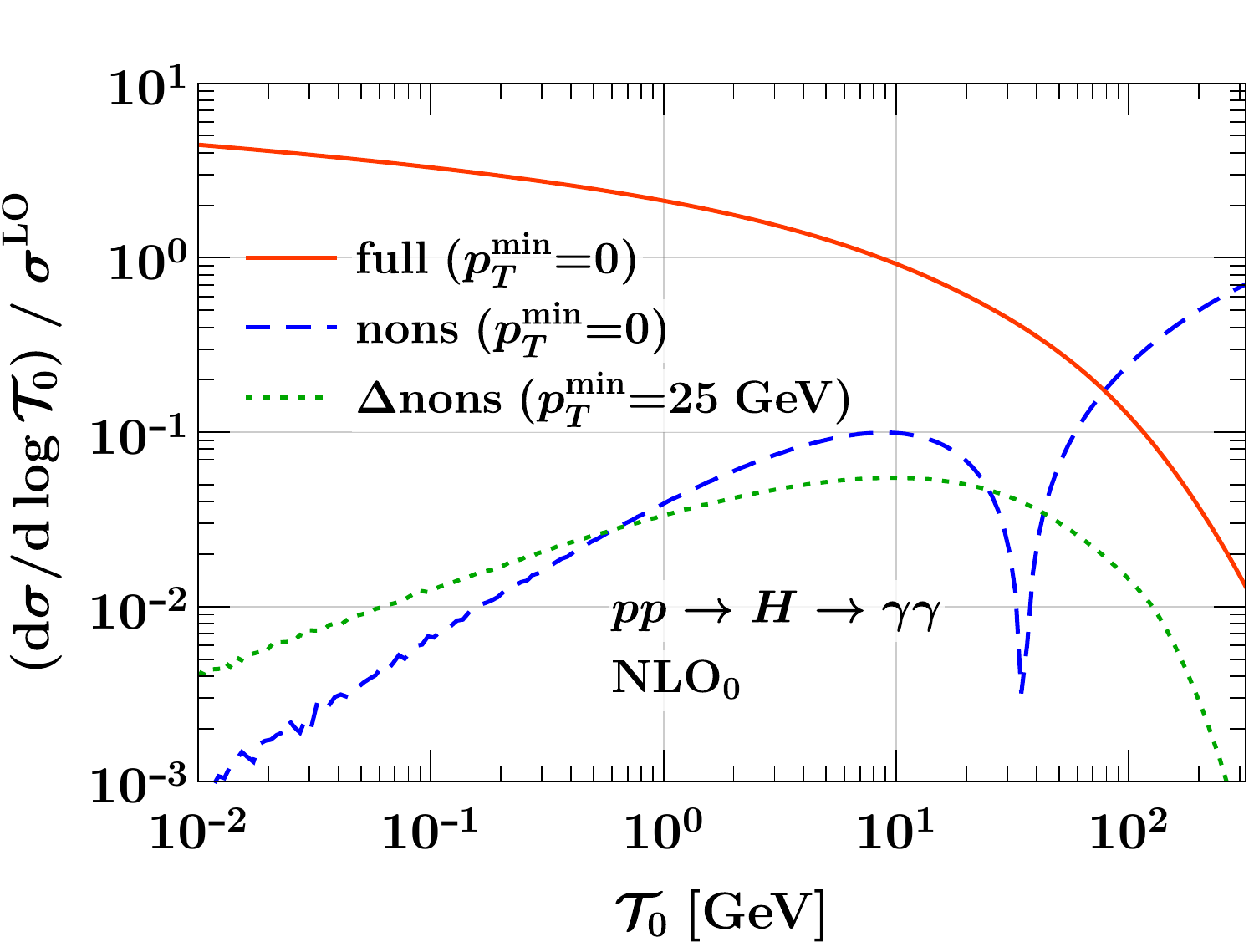}
 \caption{Comparison of Higgs production with and without a cut on the photon transverse momenta
 for the $q_T$ spectrum (left) and the $\Tau_0$ spectrum (right).}
 \label{fig:pTmin}
\end{figure}

In \fig{pTmin}, we show the $q_T$ spectrum (left) and $\Tau_0$ spectrum (right).
The red solid curve shows the full spectrum $\hat\sigma^{\rm full}$ for reference.
The blue dashed curve shows the nonsingular spectrum $\hat\sigma^{\rm nons}$
without the $p_T^\min$ cut.
Its slope shows the $\cO(q_T^2)$ and $\cO(\Tau_0)$ suppression
of the nonsingular corrections without any cuts.
For $\Tau_0$, the nonsingular terms change sign around
$\Tau_0 \approx 30\GeV$, which due to the logarithmic scale leads to
the kink of the blue-dashed curve.
The green dotted curve shows the additional nonsingular corrections $\Delta\hat\sigma^{\rm nons}$
from applying the $p_T^\min$ cut on the photons.
Its less steep slope shows the $\cO(q_T)$ and $\cO(\sqrt{\Tau_0})$ scaling,
consistent with the result of \sec{pc_acceptance}.
The cut-induced corrections dominate
up to rather large values $q_T \lesssim 5~\GeV$ and $\Tau_0 \lesssim 1~\GeV$,
and hence have a significant impact for both subtractions and resummation applications.
At typical subtraction cutoffs $q_T \lesssim 1~\GeV$
the cut-induced corrections are almost an order of magnitude enhanced, while
for $\Tau_0 \lesssim 0.1\GeV$ they are enhanced by a factor of two.

%===============================================================================
\FloatBarrier
\subsection{Photon isolation cuts}
\label{sec:numerics_pc_isolation}
%===============================================================================

\begin{figure}
 \centering
 \includegraphics[width=0.48\textwidth]{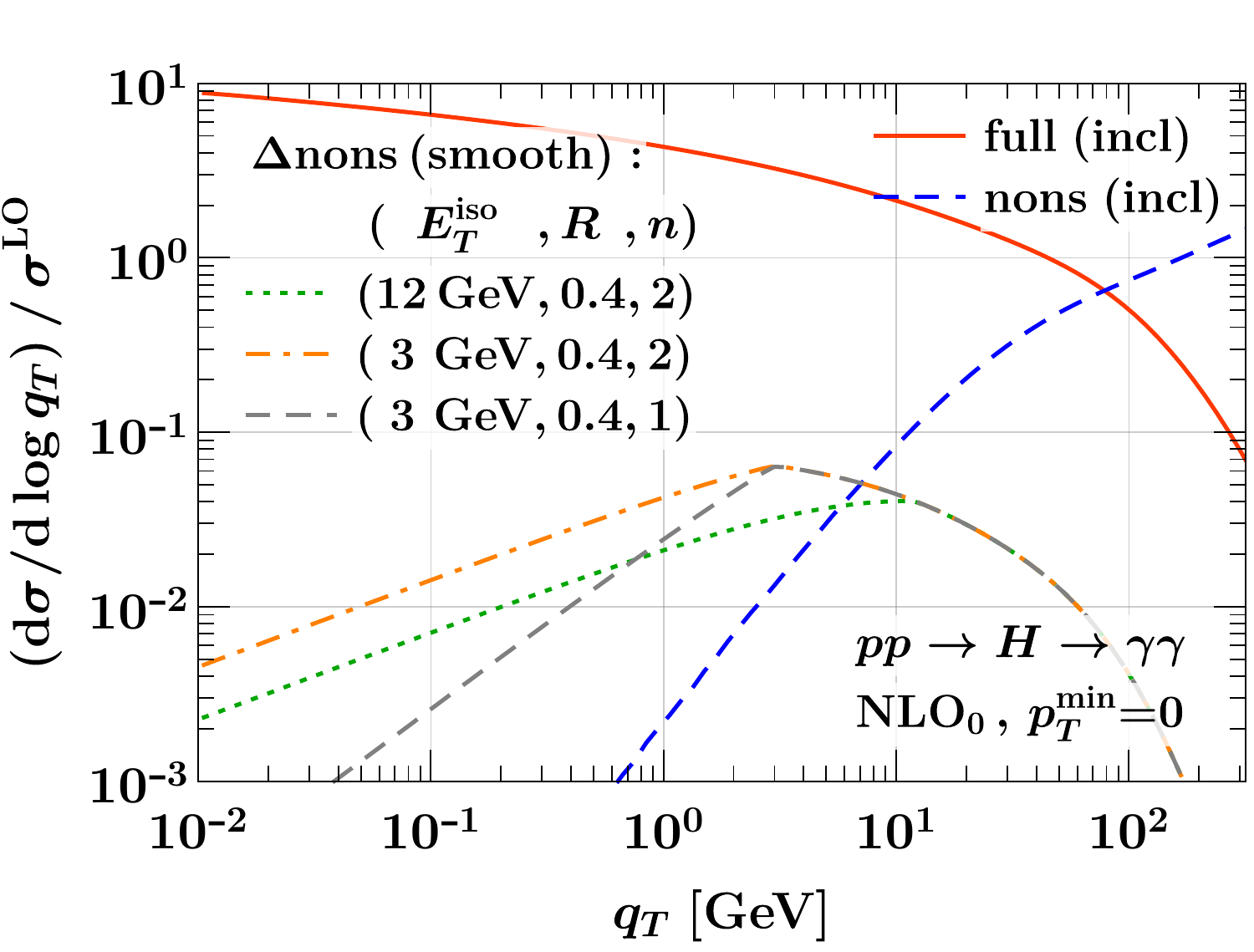}
 \quad
 \includegraphics[width=0.48\textwidth]{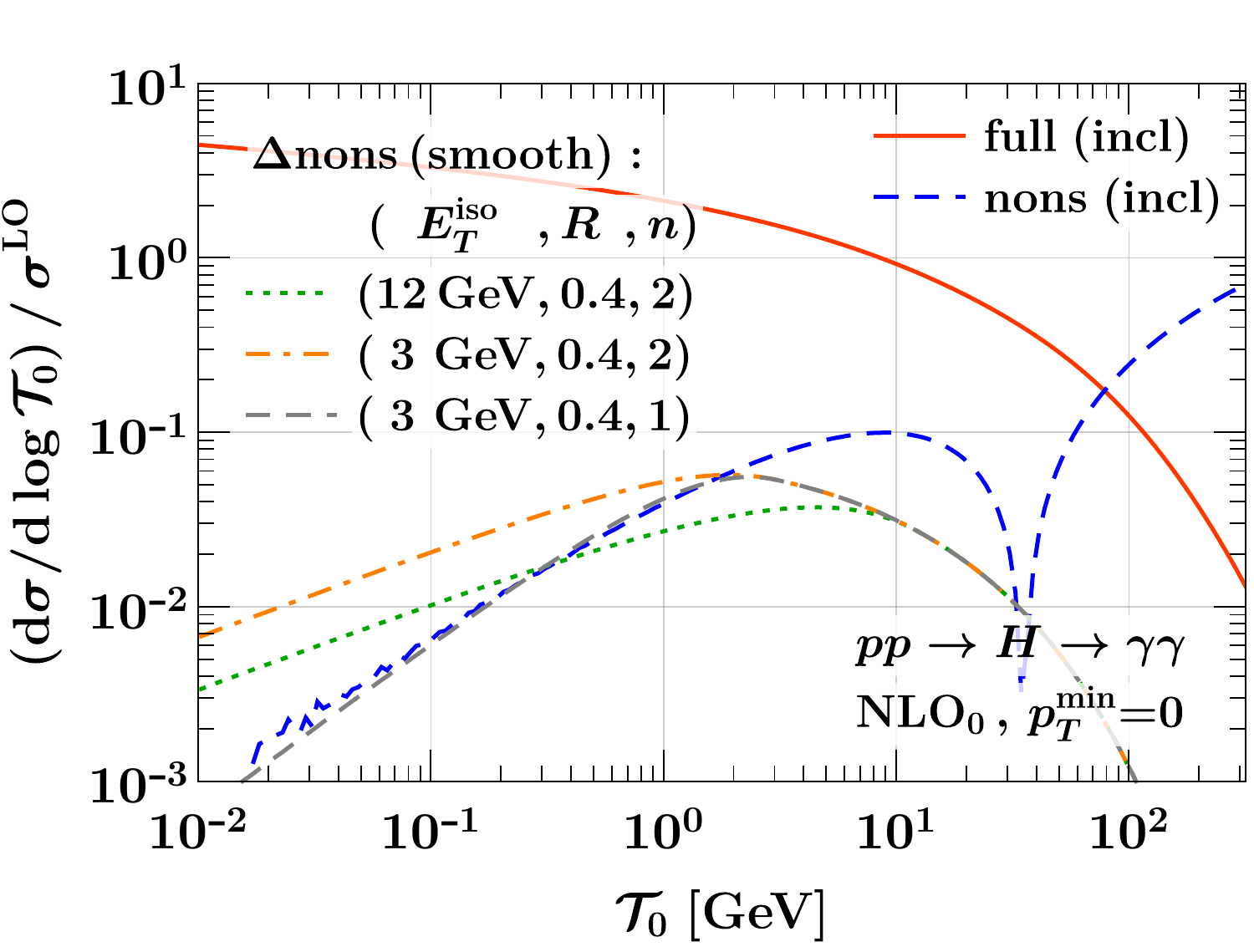}
 \caption{Power corrections in $H\to\gamgam$ with smooth-cone isolation
 for the $q_T$ spectrum (left) and the $\Tau_0$ spectrum (right).
 The red and blue dashed lines show the full and nonsingular results without isolation.
 The other curves show the additional nonsingular corrections induced by the isolation
 for different isolation parameters.}
 \label{fig:Frixione_params}
\end{figure}

\begin{figure}
 \centering
 \begin{subfigure}{\textwidth}
  \includegraphics[width=0.48\textwidth]{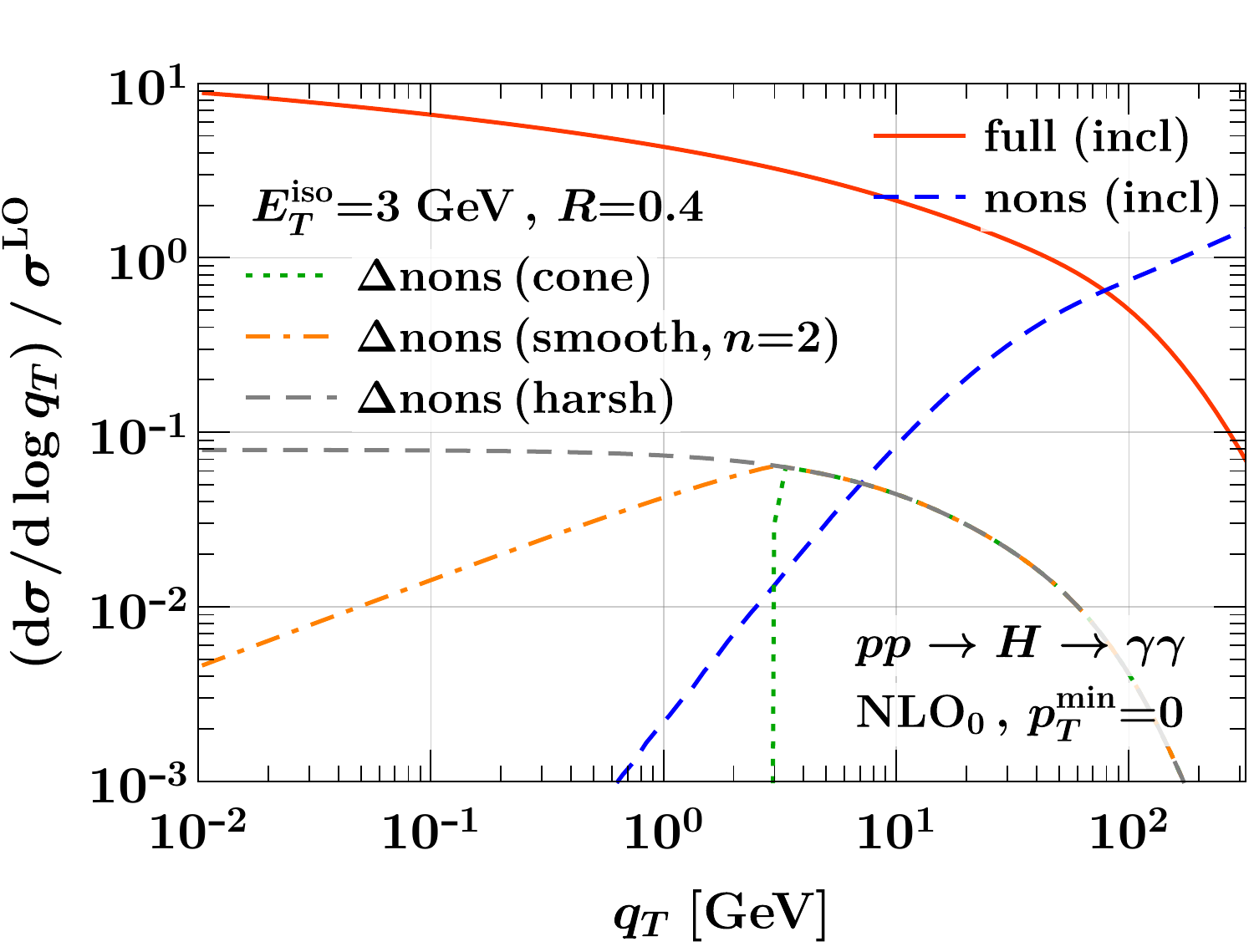}
   \quad
  \includegraphics[width=0.48\textwidth]{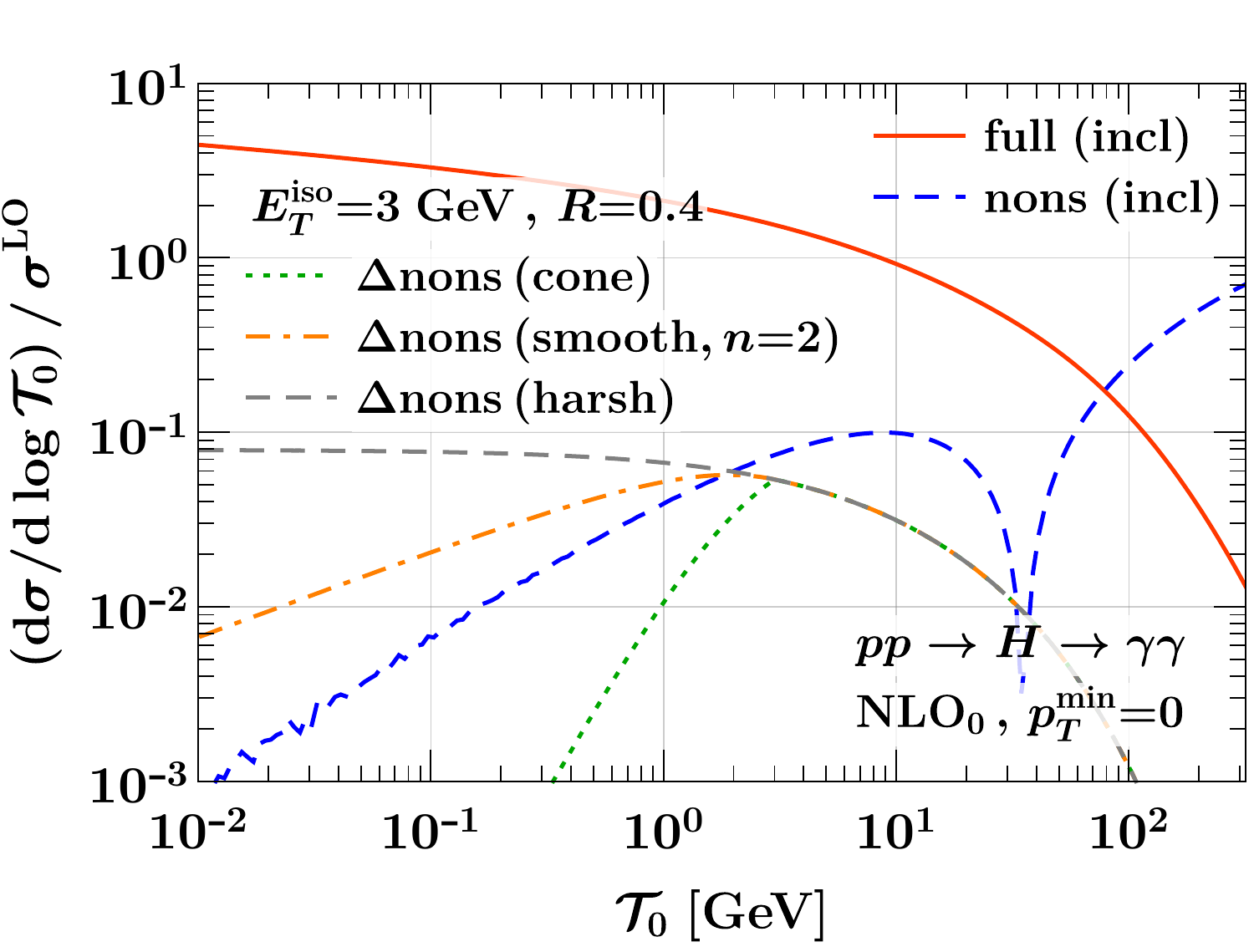}
  %%%
  \caption{$pp\to H \to \gamgam$ with $p_T^\min=0$.}
 \end{subfigure}
%%%
 \begin{subfigure}{\textwidth}
  \includegraphics[width=0.48\textwidth]{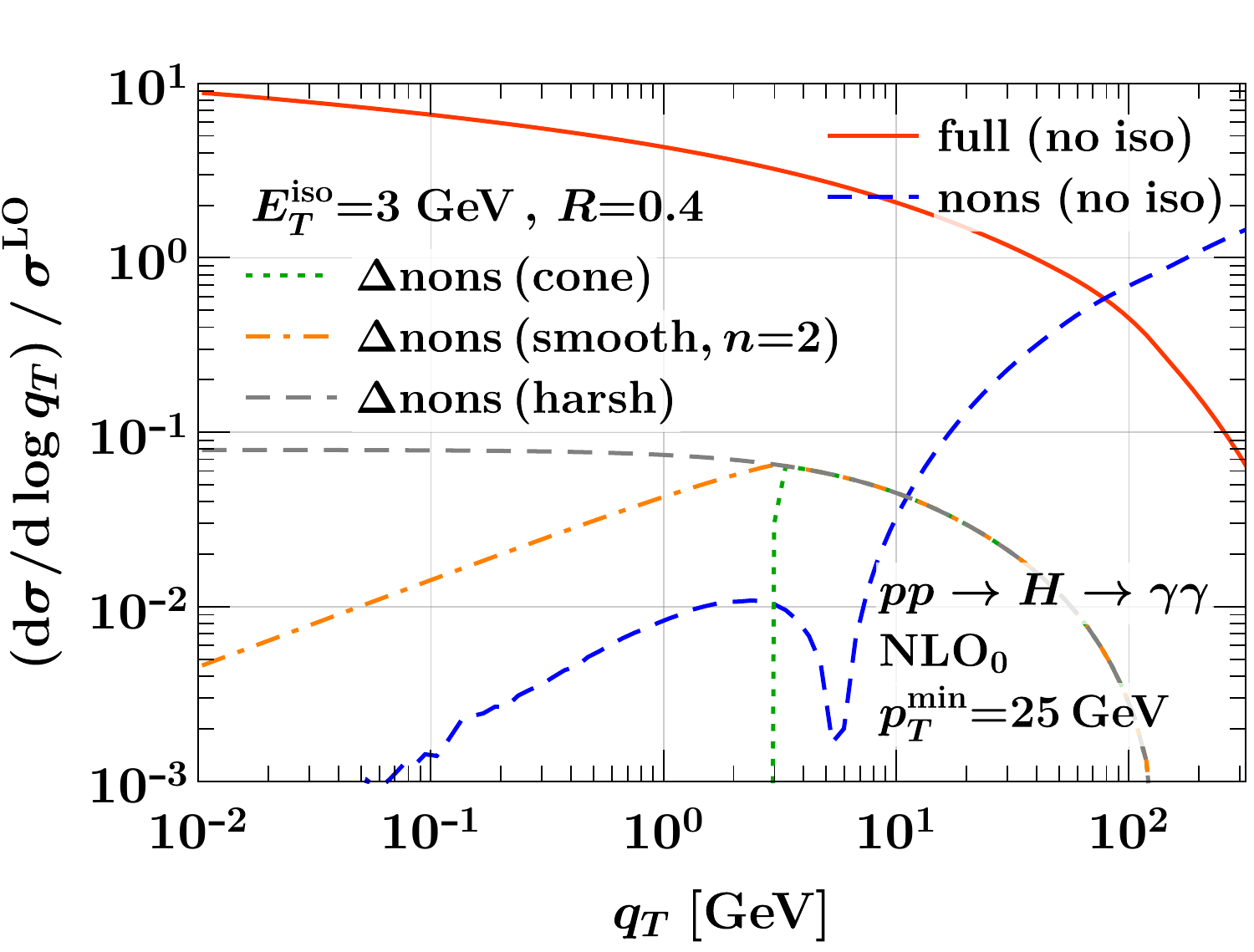}
   \quad
  \includegraphics[width=0.48\textwidth]{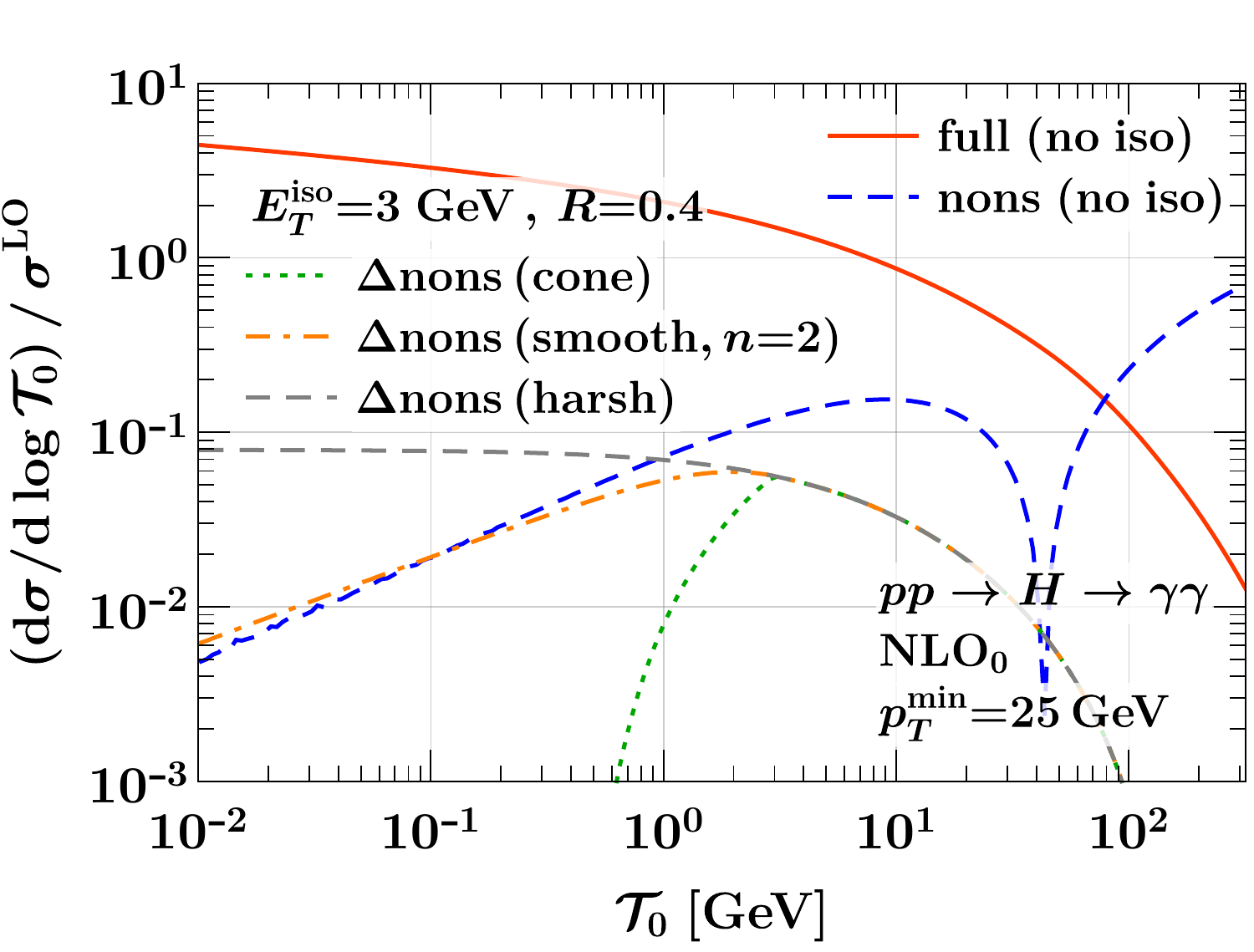}
  %%%
  \caption{$pp\to H \to \gamgam$ with $p_T^\min=25~\GeV$.}
 \end{subfigure}
%%%
 \begin{subfigure}{\textwidth}
  \includegraphics[width=0.48\textwidth]{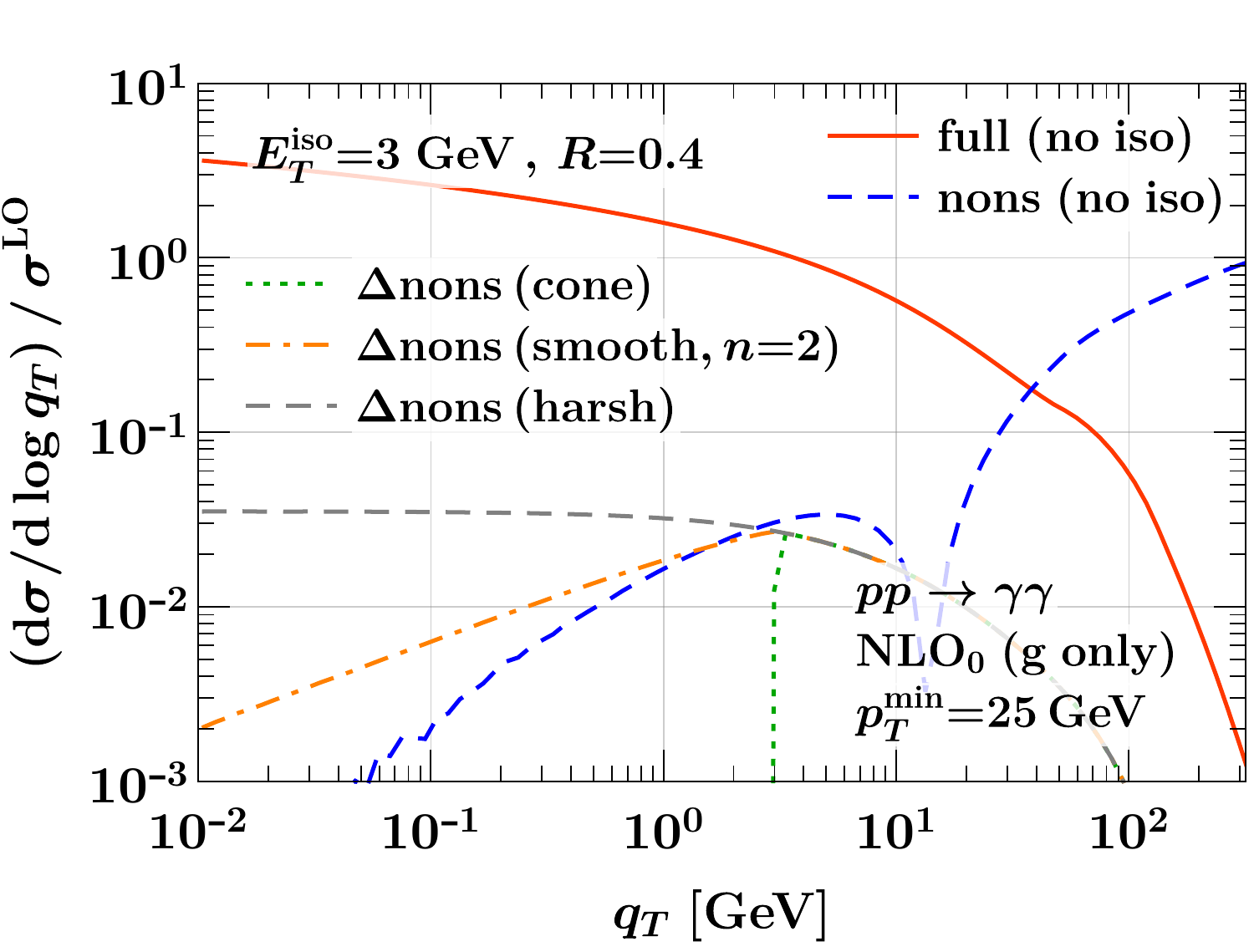}
   \quad
  \includegraphics[width=0.48\textwidth]{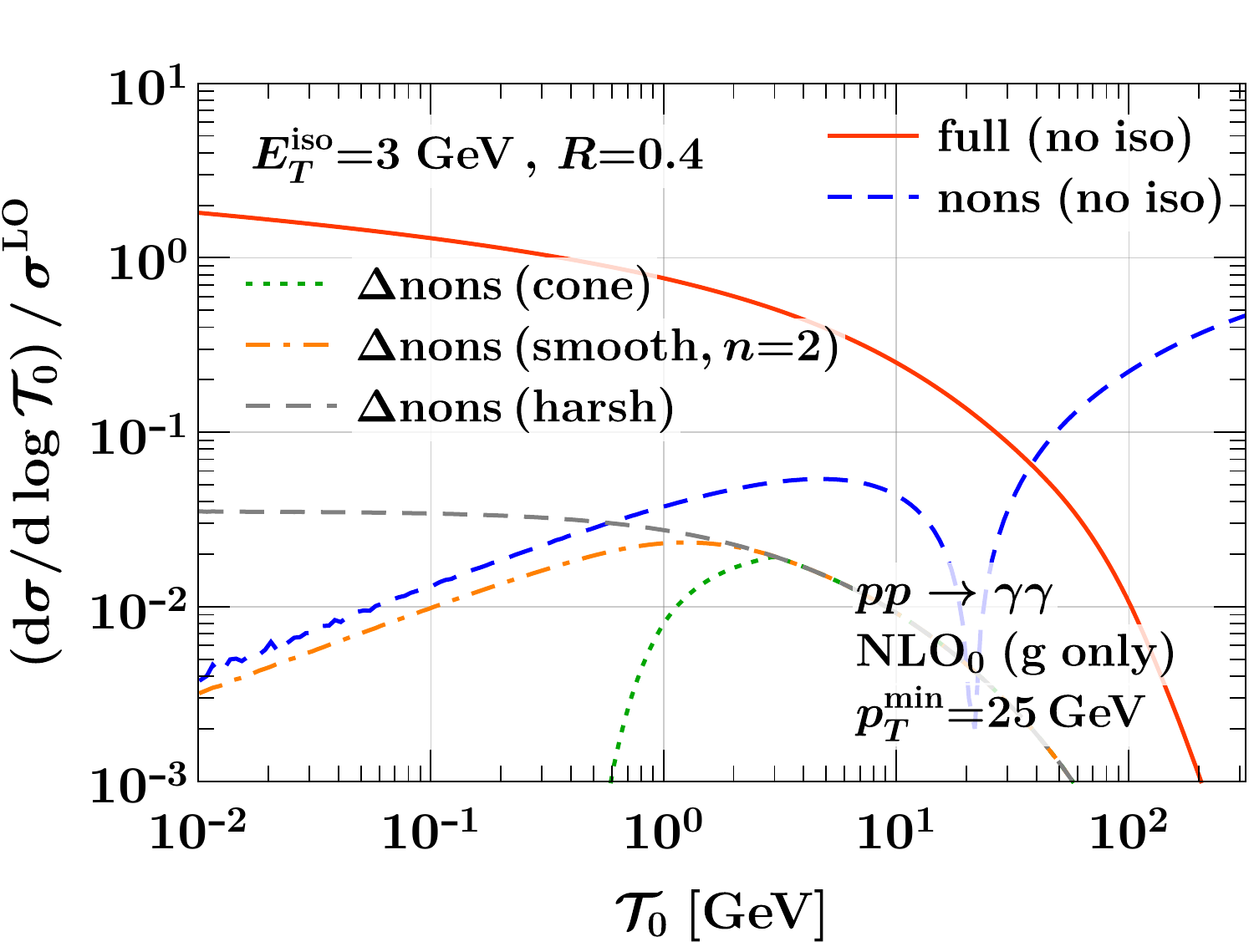}
  %%%
  \caption{$q\bar{q}\to\gamgam$ with $p_T^\min=25~\GeV$ and $m_{\gamgam} \in [120,130]~\GeV$.}
 \end{subfigure}
%%%
 \caption{Comparison of the power corrections for the $q_T$ spectrum (left) and
 the $\Tau_0$ spectrum (right) for different photon isolation methods.
 The red and blue curves show the full result and nonsingular corrections without
 any isolation. The other curves show the additional nonsingular corrections induced
 by the isolation using fixed-cone isolation (green), smooth-cone isolation (orange),
 and harsh isolation (gray).}
 \label{fig:sing_nonsing}
\end{figure}

Next, we consider the effect of photon isolation cuts.
We begin by illustrating the dependence of the power corrections for smooth-cone
isolation on the isolation parameters, as given in \eqs{pc_smooth_qT}{pc_smooth_Tau0}.
To not mix effects from the photon isolation and kinematic acceptance cuts,
we restrict ourselves to Higgs production with $p_T^\min=0$.
Since the induced power corrections depend trivially on the isolation radius $R$,
$\Delta\hat\sigma \sim R^2$, we fix $R = 0.4$ and only vary the isolation energy $E_T^\iso$
and the parameter $n$. We consider the three choices
\begin{alignat}{2} \label{eq:frix_params}
 &\text{green dotted:} \qquad &&E_T^\iso = 12~\GeV,~ R=0.4,~ n=2
\,,\nn\\
 &\text{orange dot-dashed:} \qquad &&E_T^\iso = 3~\GeV,~\,~ R=0.4,~ n=2
\,,\nn\\
 &\text{gray dashed:} \qquad &&E_T^\iso = 3~\GeV,~\,~ R=0.4,~ n=1
\,,\end{alignat}
%%%
for which we show in \fig{Frixione_params} the $q_T$ and $\Tau_0$ spectra.
The red solid curve shows the full result $\hat\sigma^{\rm full}$ for reference.
The blue dashed curve shows the nonsingular corrections $\hat\sigma^{\rm nons}$ without
isolation cuts. Its slope shows the normal $\cO(q_T^2)$ and $\cO(\Tau_0)$ suppression
(and similar to \fig{pTmin} the kink around $\Tau_0 \approx 30\GeV$ is due to a sign change).
The additional curves as stated in \eq{frix_params} show the additional
nonsingular correction $\Delta\hat\sigma^{\rm nons}$ from the different isolations requirements,
which for small $q_T$ and $\Tau_0$ obey the $\cO(q_T^{1/n})$ and $\cO(\Tau_0^{1/n})$
behavior as predicted by \eqs{pc_smooth_qT}{pc_smooth_Tau0}.
The gap between the green-dotted and orange-dot-dashed curves corresponds to a factor of $2$, correctly reflecting the scaling of the power corrections with $\sqrt{\ETiso}$ for $n=2$.
Above $q_T \ge \ETiso$ and $\Tau_0 \ge \ETiso$, the different isolations agree
as in this limit each emission that falls into an isolation cone is necessarily too energetic
to be allowed, independently of the chosen isolation method. In this region, the isolation
is in fact a leading-power effect, while below this region it becomes a power correction
which leads to the kink at $q_T = \ETiso$ and $\Tau_0 = \ETiso$.
(For $\Tau_0$, this follows from the explicit calculation presented in \sec{cone_isolation}.)

Overall, we find that in each case the smooth-cone isolation yields large
additional corrections, which as expected from the relative scaling are significantly
enhanced compared to the normal power corrections (blue dashed), and which exhibit
a very slow convergence to zero for $q_T \to 0$ or $\Tau_0 \to 0$. The relative
enhancement is particularly severe for $q_T$, easily exceeding an order of magnitude for
$q_T \lesssim 1\GeV$.
This suggests that calculations of processes involving smooth-cone isolation
with $q_T$ or $\Tau_N$ subtractions should prefer a loose isolation, which however
goes opposite to the recommendation of \refscite{Andersen:2014efa,Badger:2016bpw,Catani:2018krb}
to employ tight cuts in order for smooth-cone isolation to yield similar results as fixed-cone isolation.

In \fig{sing_nonsing}, we compare fixed-cone, smooth-cone, and harsh isolations.
The top (middle) row shows Higgs production in the diphoton decay mode
with a cut $p_T^\min = 0$ ($p_T^\min=25~\GeV$) on the photons.
The bottom row shows direct diphoton production $pp \to \gamgam$ with $p_T^\min=25~\GeV$,
where only the $q\bar q\to \gamma\gamma g$ channel is taken into account to avoid fragmentation contributions.
In all figures, the red solid curves show the full result $\hat\sigma^{\rm full}$
for reference. The blue dashed curves show the nonsingular corrections $\hat\sigma^{\rm nons}$
without any isolation but including the $p_T^\min$ cut.
The additional nonsingular corrections induced by the isolation
are shown in green dotted for fixed-cone isolation, orange dot-dashed for smooth-cone isolation with $n=2$,
and in gray dashed for harsh isolation. In each case, we use $R=0.4$ and $\ETiso = 3~\GeV$.

For the $q_T$ spectrum, we see that cone isolation has no power corrections for $q_T \le \ETiso$,
and likewise almost negligible corrections to the $\Tau_0$ spectrum for $\Tau_0 \le \ETiso$,
consistent with our findings in \sec{pc_isolation}.
In contrast, smooth-cone isolation shows the predicted much weaker suppression
of $\cO(q_T^{1/n})$ and $\cO(\Tau_0^{1/n})$.
As a result, it yields in all cases sizable additional power corrections,
which for $q_T$ clearly dominate over the corrections without isolation,
both with and without the $p_T^\min$ cut. For $\Tau_0$, they are of the same order
as the corrections induced by the $p_T^\min$ cut, while for $p_T^\min=0$ the isolation
again dominates over the inclusive nonsingular corrections.
Finally, the harsh isolation yields an almost constant correction on the logarithmic plot,
which translates into leading-power correction in $1/q_T$ and $1/\Tau_0$.
Note that these are not integrable as $q_T,\Tau_0\to0$, reflecting the factorization violation
from the infrared-unsafe isolation procedure.

%%%%%%%%%%%%%%%%%%%%%%%%%%%%%%%%%%%%%%%%%%%%%%%%%%%%%%%%%%%%%%%%%%%%%%%%%%%%%%%%
\section[\texorpdfstring{$\Tau_N$}{TauN} subtractions including measurement cuts]
        {$\Tau_N$ subtractions including measurement cuts}
\label{sec:subtractions}
%%%%%%%%%%%%%%%%%%%%%%%%%%%%%%%%%%%%%%%%%%%%%%%%%%%%%%%%%%%%%%%%%%%%%%%%%%%%%%%%

In this section, we discuss how all cut-induced power corrections can be
accounted for exactly in the subtraction procedure. Our starting point are
differential $\Tau_N$ subtractions~\cite{Gaunt:2015pea}, using which
the cross section with a measurement $X$ is given by
%%%
\begin{align} \label{eq:tau_subtraction}
\sigma(X)
&= \sigma^\sub(X, \tau_\off)
+ \int\!\df\tau\, \biggl[ \frac{\df\sigma(X)}{\df\tau}
- \frac{\df\sigma^\sub(X)}{\df\tau} \theta(\tau < \tau_\off) \biggr]
\nn \\
&= \sigma^\sub(X, \tau_\off)
+ \int^{\tau_\off}\!\df\tau\, \biggl[ \frac{\df\sigma(X)}{\df\tau}
- \frac{\df\sigma^\sub(X)}{\df\tau} \biggr]
+ \int_{\tau_\off} \!\df\tau\, \frac{\df\sigma(X)}{\df\tau}
\,.\end{align}
%%%
As in \sec{review_subtractions},
$\tau$ stands for any (dimensionless) $N$-jet resolution variable for which a LP
factorization theorem is known.
The differential subtraction term $\df\sigma^\sub(X)/\df\tau$
captures the leading-power singularities for $\tau\to 0$, which means it satisfies
%%%
\begin{align}
\frac{\df\sigma^\sub(X)}{\df\tau} = \frac{\df\sigma^{(0)}(X)}{\df\tau}
\bigl[1 + \ord{\tau}\bigr]
\,,\end{align}
%%%
such that the integrand in square brackets in \eq{tau_subtraction} is a power
correction with at most integrable singularities for $\tau\to 0$, and so the integral
can be carried out numerically. Since the integral exists and is finite, the point $\tau = 0$
is irrelevant, which means the integrand is never evaluated at $\tau = 0$. Hence,
the full result for $\df\sigma(X)/\df\tau$ is only needed for nonzero $\tau > 0$
and thus reduces to performing the NLO Born+1-parton calculation. Similarly the
distributional structure of $\df\sigma^\sub(X)/\df\tau$ at $\tau = 0$ is not needed for the differential
subtraction terms, which are fully known to N$^3$LO for both $q_T$ and $\Tau_0$ subtractions~\cite{Billis:2019vxg}.
The first term in \eq{tau_subtraction}
is the cumulant of $\df\sigma^\sub(X)/\df\tau$ up to $\tau_\off$. Its evaluation
does require the full distributional structure of $\df\sigma^\sub(X)/\df\tau$.

Note that in principle the integrand does need to be sampled arbitrarily close to
$\tau = 0$, but due to the subtraction the contribution from a region $\tau < \delta$
is of $\ord{\delta}$.
This is similar to the fact that even in a fully local subtraction method the
real-emission phase-space formally needs to be sampled arbitrarily close to the IR-singular
region, but the subtractions ensure that the total subtracted integrand
is well-behaved, so the contribution from a region of size $\delta$ around the singularity
only contributes an amount of $\ord{\delta}$. Letting $\delta\to 0$ still requires
evaluating the real-emissions matrix elements arbitrarily close to the singularity,
and to avoid numerical instabilities due to arbitrarily large numerical cancellations
one always has a technical cutoff $\delta$ that cuts out the actual singular points
of phase space.

The parameter $\tau_\off$ determines the range over which the subtractions
act, and by taking  $\tau_\off \sim 1$ there are no large numerical cancellations
between the first and second term in \eq{tau_subtraction}.
(In the context of resummation, $\tau_\off$ corresponds to where the $\tau$
resummation is turned off.)
The slicing method described in \sec{review_subtractions} is obtained from
\eq{tau_subtraction} by taking $\tau_\off = \tau_\cut$, see \eq{nsub_master}.
In this case, the integral below $\tau_\off = \tau_\cut$
corresponds to $\Delta\sigma(X, \tau_\cut)$ and is neglected, which induces the
power corrections.
In contrast, \eq{tau_subtraction} is exact
and involves no neglected power corrections.

The practical challenge in implementing \eq{tau_subtraction} is that the NLO
calculation for $\df\sigma(X)/\df\tau$ has to be obtained as a function of
$\tau$. In general this is not easy as it requires to organize the
integration over the real-emission phase space in such a way that $\tau$
is preserved, which by default is not the case for standard NLO subtractions.
For a more detailed discussion we refer to \refcite{Gaunt:2015pea}.

To make the differential subtractions more viable in practice, we can follow
the same basic strategy as in \sec{setup_NLO} to separate the different sources
of power corrections.
We first note that the LP singular contribution only depends
on the Born phase space. That is, the factorization theorem for $\tau$ is
always fully differential in the Born phase space, which involves choosing a
specific set of kinematic variables to parametrize the Born phase space.
The measurement $X$ is then evaluated on this reference Born phase space.
In other words, constructing
$\df\sigma^\zero(X)/\df\tau$ involves choosing a Born projection $\hat\Phi_N(\Phi_{N+k})$
from the real-emission phase-space with $k$ additional emissions, $\Phi_{N+k}$,
to the Born phase space, $\Phi_N$.
For color-singlet production ($N = 0$), a typical choice is to use $Q$ and $Y$
as the Born variables, as we did in \sec{setup_NLO} above.
The LP measurement function that actually enters
in $\df\sigma^\zero(X)/\df\tau$ is then given by
%%%
\begin{equation}
f_X^\zero(\Phi_{N+k}) = f_X[\hat\Phi_N(\Phi_{N+k})]
\,.\end{equation}
%%%
For color-singlet production at NLO, this is precisely the LP
term on the right-hand side of \eq{fX_LP}.
Denoting this LP measurement by $X^\zero$, we therefore have
%%%
\begin{equation}
\frac{\df\sigma^\sub(X)}{\df\tau} \equiv \frac{\df\sigma^\sub(X^\zero)}{\df\tau}
\,, \qquad
\sigma^\sub(X, \tau_\off) \equiv \sigma^\sub(X^\zero, \tau_\off)
\,.\end{equation}
%%%

Next, we can consider the full cross section but with the measurement replaced by
this LP Born reference measurement, $\df\sigma(X^\zero)/\df\tau$.
By adding and subtracting it, we can rewrite \eq{tau_subtraction} as
%%%
\begin{align} \label{eq:tau_subtraction_withref}
\sigma(X)
&= \sigma^\sub(X^\zero, \tau_\off)
+ \int\!\df\tau\, \biggl[ \frac{\df\sigma(X^\zero)}{\df\tau}
- \frac{\df\sigma^\sub(X^\zero)}{\df\tau} \theta(\tau < \tau_\off) \biggr]
+ \int\!\df\tau\, \frac{\df\sigma(X-X^\zero)}{\df\tau}
\nn \\
&\equiv \sigma(X^\zero) + \sigma(X - X^\zero)
\,.\end{align}
%%%
We have now isolated the two different sources of power corrections.
The sum of the first two terms in the first line of \eq{tau_subtraction_withref}
is the calculation of the
reference cross section $\sigma(X^\zero)$ using differential $\tau$ subtractions.
Since it involves the same reference measurement $X^\zero$ everywhere,
the difference $\df\sigma(X^\zero) - \df\sigma^\sub(X^\zero)$ does not
involve any cut-induced power corrections, hence reducing the problem
of power corrections to the normal and well-studied case, and for which
the power corrections can be systematically calculated if necessary~\cite{Moult:2016fqy,
Boughezal:2016zws, Moult:2017jsg, Boughezal:2018mvf, Ebert:2018lzn, Ebert:2018gsn, Boughezal:2019ggi}.
In particular, if the implementation of the differential $\tau$ subtractions
proves too difficult in practice, this contribution could be calculated with the
slicing approach (see below).

The last term in \eq{tau_subtraction_withref} amounts to measuring the difference
between $X$ and $X^\zero$ on the full cross section. Here we exploited that the
difference of the two cross sections can be combined into a single cross section,
as the only difference lies in the measurement,
%%%
\begin{equation} \label{eq:delta_sigma}
\int\!\df\tau\biggl[\frac{\df\sigma(X)}{\df\tau} - \frac{\df\sigma(X^\zero)}{\df\tau}\biggr]
= \int\!\df\tau\, \frac{\df\sigma(X-X^\zero)}{\df\tau}
= \sigma(X - X^\zero)
\,.\end{equation}
%%%
That is, $\sigma(X-X^\zero)$
contains the difference of the full and LP measurement functions,
$f_X(\Phi_{N+k}) - f_X^\zero(\Phi_{N+k})$. For example, for color-singlet
production at NLO, $\df\sigma(X - X^\zero)$ is precisely given by \eq{sigma_iso1}.
Since for any infrared safe $X$ this measurement difference vanishes in the singular
limit, $\sigma(X - X^\zero)$ still amounts to effectively performing a
Born+1-parton calculation at one lower order.
It contains all cut-induced power corrections, which as we discussed can be
potentially large, and it should therefore be treated exactly.
Since it can be formulated as a specific choice
of measurement, it can be implemented straightforwardly into
existing NLO calculations. Once this is done, the explicit dependence
on $\tau$ disappears. (In general it might still be implicit through the choice of $X^\zero$.)
One might say that the reference cross section $\df\sigma(X^\zero)$ in \eq{delta_sigma}
effectively acts as a fully local subtraction term. However, this is somewhat misleading,
since the IR singularities do not cancel in the difference of two singular contributions.
Rather, they are simply regulated by performing an IR-safe Born+1-parton
measurement.

When performing the calculation of $\sigma(X-X^\zero)$ one might still have
to integrate near the singular region of phase space, but only to the extent
to which the full measurement is sensitive to, which is the best one can hope for.
For example, if $X$ contains isolation cuts, then $X^\zero$ will contain no
isolation cuts. The difference $X - X^\zero$ then measures the cross section
that is removed by the isolation, which is sensitive to real emissions with energies
down to $\ETiso$, while below that the difference of the two measurements explicitly vanishes.
For selection cuts, one can still get sensitive to arbitrarily soft emissions,
e.g., when measuring the $p_T$ of the photons in $H\to\gamma\gamma$ very close to the
Born limit $p_T = m_H/2$. However, this is a well-known feature of such cuts
and inherent to the measurement itself and not related the subtraction method.

From the above discussion, we can also see the connection to the projection-to-Born
method~\cite{Cacciari:2015jma}. It amounts to the special case where the reference cross
section $\df\sigma(X^\zero)$ is known analytically or from some other calculation,
while the last term
is precisely the effective Born+1-parton calculation that also appears in the
projection-to-Born method. In other words, the projection-to-Born method is
simply the statement that $\sigma(X)$ can be calculated as
%%%
\begin{equation}
\sigma(X) = \sigma(X^\zero) + \sigma(X - X^\zero)
\,,\end{equation}
%%%
when the full cross section $\sigma(X^\zero)$ for some reference
measurement $X^\zero$ is already known, and the correction term
$\sigma(X - X^\zero)$ is calculated
by evaluating the $X-X^\zero$ measurement
for the lower-order Born+1-parton calculation as described above.

To conclude, we note that if the reference cross section $\sigma(X^\zero)$ is
obtained via a global $\tau$ slicing, one can of course combine both Born+1-parton
calculations into a single one,
%%%
\begin{align} \label{eq:tau_subtraction_slicing_withref}
\sigma(X)
&= \sigma^\sub(X^\zero, \tau_\cut) + \sigma[X - X^\zero\theta(\tau<\tau_\cut) ]
+ \Delta\sigma(X^\zero, \tau_\cut)
\,.\end{align}
%%%
This makes it explicit that in contrast to \eq{nsub_master}, here the power corrections
$\Delta\sigma(X^\zero, \tau_\cut)$ are only those for the chosen reference measurement. The cut-induced power
corrections are accounted for by the Born+1-parton calculation in the second term, because
it correctly captures the difference $X - X^\zero$ below $\tau_\cut$.

%%%%%%%%%%%%%%%%%%%%%%%%%%%%%%%%%%%%%%%%%%%%%%%%%%%%%%%%%%%%%%%%%%%%%%%%%%%%%%%%
\FloatBarrier
\section{Conclusions}
\label{sec:conclusions}
%%%%%%%%%%%%%%%%%%%%%%%%%%%%%%%%%%%%%%%%%%%%%%%%%%%%%%%%%%%%%%%%%%%%%%%%%%%%%%%%

We have studied the impact of kinematic selection cuts
and isolation requirements for leptons and photons on the $q_T$ and $N$-jettiness
subtraction methods. Using a simplified one-loop calculation, we analytically determined
the scaling of power corrections induced by these cuts including their dependence
on the isolation method and its parameters.
We find that both selection cuts and isolation induce additional power corrections
that are parametrically enhanced relative to the usual, cut-independent power corrections
inherent to the $q_T$ and $\Tau_0$ factorization theorems.
We have also discussed how the cut effects can be fully incorporated into the subtraction,
thereby avoiding the additional power corrections, by employing
differential subtractions for them instead of a global slicing method.

To summarize our key findings, we expand the differential $q_T$ and $\Tau_0$ spectra as
\begin{align} \label{eq:summary}
 \frac{\df\sigma(X)}{\df Q^2 \df Y \df q_T^2} &
 = \frac{\df\sigma^{(0)}(X)}{\df Q^2 \df Y \df q_T^2}
   \times \Bigl[1 + \cO\bigl[(q_T^2/Q^2)^{m}\bigr] \Bigr]
\,,\nn\\
 \frac{\df\sigma(X)}{\df Q^2 \df Y \df \Tau_0} &
 = \frac{\df\sigma^{(0)}(X)}{\df Q^2 \df Y \df \Tau_0}
   \times \Bigl[1 + \cO\bigl[(\Tau_0/Q)^{m}\bigr] \Bigr]
\,,\end{align}
where $\sigma^{(0)}$ are the leading-power limits predicted by the factorization theorems.
We find the following power corrections
in the square brackets in \eq{summary} for typical selection and isolation cuts:
\begin{itemize}
 \item For inclusive processes without any cuts, one has $m=1$.
 \item A typical $p_T > p_T^\min$ selection cut for photons or leptons yields enhanced power
       corrections with $m=1/2$ and proportional to $\sim p_T^\min/Q$.
       Since this arises from breaking azimuthal symmetry that is only present in the
       Born process, we expect a similar enhancement for generic fiducial cuts.
 \item All photon isolation methods yield \emph{leading-power} corrections ($m=0$)
       for $q_T > \ETiso$ and $\Tau_0 > \ETiso$, respectively,
       which are proportional to the size of the isolation cone $\sim \cO(R^2)$.
 \item At one loop, fixed-cone isolation induces no corrections for $q_T < \ETiso$
       and highly suppressed corrections ($m=2$) for $\Tau_0 < \ETiso$. At higher
       orders one can expect nontrivial corrections also below $\ETiso$, which
       should be power suppressed.
 \item Smooth-cone isolation as defined in \eq{chi2} yields power corrections
       scaling as $m=1/(2n)$ for $q_T$ and $m=1/n$ for $\Tau_0$, respectively.
       They are further enhanced by an overall factor $(Q/\ETiso)^{1/n}$.
\end{itemize}
In general, tight cuts can thus yield significantly enhanced power corrections.
The enhancement is most severe for smooth-cone isolation with $q_T$ subtractions.
We have numerically verified and studied these findings for the examples of
$pp \to H \to \gamgam$ and $pp \to \gamgam$.

While our analysis is based on an explicit one-loop study, we expect the
dominant qualitative behavior to persist at NNLO and beyond, since the same
kinematic effects will also appear at higher orders.
For example, our results immediately apply to real-virtual contributions
at higher orders involving a single real emission.
For contributions with two or more real emissions additional nontrivial
kinematic correlations among multiple emissions are likely to lead to
additional effects, e.g., one can expect the kinks at
$q_T = \ETiso$ and $\Tau_0 = \ETiso$ to get smeared out.
It seems extremely unlikely though that such effects from multiple emissions
could somehow improve the behavior that is already present for a single real
emission -- one might hope that they do not make things worse.
Note that at order $\as^n$, the inclusive power corrections contain up to $2n-1$
logarithms $\ln(Q/q_T)$ and $\ln(Q/\Tau_0)$, respectively, and it would be
interesting to study in detail to what extent the enhanced power corrections
also receive such additional logarithmic factors, which would make them
numerically even more important.

Our results provide an important step for a better understanding of power corrections
whenever kinematic selection cuts or isolation cuts are applied.
This is crucial both for subtraction methods and the resummation of large logarithms in such processes.
In principle, our technique can be employed to exactly calculate the induced corrections.
In practice, it will however be more advantageous to account for all cut-induced corrections
within the subtraction method itself as discussed in \sec{subtractions}.

%%%%%%%%%%%%%%%%%%%%%%%%%%%%%%%%%%%%%%%%%%%%%%%%%%%%%%%%%%%%%%%%%%%%%%%%%%%%%%%%
\begin{acknowledgments}
We thank Ian Moult for earlier collaboration on parts of this work.
We also thank Johannes Michel, Iain Stewart, and Kerstin Tackmann for helpful
discussions.
This work was supported in part by the Office of Nuclear Physics of the U.S.\
Department of Energy under Contract No.\ DE-SC0011090,
the Alexander von Humboldt Foundation through a Feodor Lynen Research Fellowship,
the Deutsche Forschungsgemeinschaft (DFG) under Germany's Excellence
Strategy -- EXC 2121 ``Quantum Universe'' -- 390833306,
and the PIER Hamburg Seed Project PHM-2019-01.
\end{acknowledgments}
%%%%%%%%%%%%%%%%%%%%%%%%%%%%%%%%%%%%%%%%%%%%%%%%%%%%%%%%%%%%%%%%%%%%%%%%%%%%%%%%

\bibliography{literature}
\bibliographystyle{jhep}

\end{document}